\documentclass[12pt]{article}
\usepackage{latexsym}
\usepackage{epsfig,amssymb,euscript,slashed}
\usepackage[linktocpage]{hyperref}
\usepackage{amsmath}
\usepackage[noadjust]{cite}
\usepackage{mathtools}
\usepackage{physics}
\usepackage{extarrows}
\usepackage{xcolor}
\usepackage{float}
\usepackage{graphicx}
\usepackage{subcaption}
\usepackage{tcolorbox}
\usepackage{MnSymbol}
\usepackage{tikz-feynman}
\usepackage{array,calc,epsfig}
\usepackage{bbm}
\usepackage{fancybox}

\newcommand{\hb}{\bar{h}}
\newcommand{\pd}{\partial}
\newcommand{\pdb}{\bar{\partial}}
\newcommand{\mb}{\bar{m}}
\newcommand{\zb}{\bar{z}}

\DeclareMathOperator{\Li}{\operatorname{Li}}

\newcommand{\poc}[2]{(#1)_{#2}}

\numberwithin{equation}{section} 
\usepackage[]{graphicx}
\usepackage{color}

\begin{document}
\font\cmss=cmss10 \font\cmsss=cmss10 at 7pt

\begin{flushright}{  
\scriptsize QMUL-PH-20-18}
\end{flushright}
\hfill
\vspace{18pt}
\begin{center}
{\Large 
\textbf{The Regge limit of AdS$_3$ holographic correlators
}}

\end{center}

\vspace{8pt}
\begin{center}
{\textsl{Stefano Giusto$^{\,a, b}$, Marcel R. R. Hughes$^{\,c}$ and Rodolfo Russo$^{\,c}$}}

\vspace{1cm}

\textit{\small ${}^a$ Dipartimento di Fisica ed Astronomia ``Galileo Galilei",  Universit\`a di Padova,\\Via Marzolo 8, 35131 Padova, Italy} \\  \vspace{6pt}

\textit{\small ${}^b$ I.N.F.N. Sezione di Padova,
Via Marzolo 8, 35131 Padova, Italy}\\
\vspace{6pt}

\textit{\small ${}^c$ Centre for Research in String Theory, School of Physics and Astronomy\\
Queen Mary University of London,
Mile End Road, London, E1 4NS,
United Kingdom}\\
\vspace{6pt}

\end{center}

\vspace{12pt}

\begin{center}
\textbf{Abstract}
\end{center}

\vspace{4pt} {\small
\noindent 
We study the Regge limit of 4-point AdS$_3 \times S^3$ correlators in the tree-level supergravity approximation and provide various explicit checks of the relation between the eikonal phase derived in the bulk picture and the anomalous dimensions of certain double-trace operators. We consider both correlators involving all light operators and HHLL correlators with two light and two heavy multi-particle states. These heavy operators have a conformal dimension proportional to the central charge and are pure states of the theory, dual to asymptotically AdS$_3 \times S^3$ regular geometries. Deviation from AdS$_3 \times S^3$ is parametrised by a scale $\mu$ and is related to the conformal dimension of the dual heavy operator. In the HHLL case, we work at leading order in $\mu$ and derive the CFT data relevant to the bootstrap relations in the Regge limit. Specifically, we show that the minimal solution to these equations relevant for the conical defect geometries is different to the solution implied by the microstate geometries dual to pure states.}

\vspace{1cm}

\thispagestyle{empty}

\vfill
\vskip 5.mm
\hrule width 5.cm
\vskip 2.mm
{
\noindent  {\scriptsize e-mails:  {\tt stefano.giusto@pd.infn.it, m.r.r.hughes@qmul.ac.uk, r.russo@qmul.ac.uk} }
}

\setcounter{footnote}{0}
\setcounter{page}{0}

\newpage

\tableofcontents


\section{Introduction}
\label{sec:intro}

The domain of high-energy, large impact parameter scattering provides an interesting laboratory in which to analyse different gravitational theories in a quantitative way. In this regime, often called the Regge limit, the $2 \to 2$ interaction can be analysed using the eikonal approximation. In the context of perturbative string theory, the study was initiated in~\cite{Amati:1987wq,Amati:1987uf} where a stringy eikonal operator was derived from four-point amplitudes (at tree and loop level) with external massless states. A complementary geometric description of the same process is in terms of a particle propagating in a shock wave background, representing the other (highly boosted) particle~\cite{'tHooft:1987rb}. The same eikonal problem was studied in the setting of AdS/CFT, starting from~\cite{Cornalba:2006xk,Cornalba:2006xm,Cornalba:2007zb}: in this case the observables playing the role of the four-point amplitudes are CFT four-point correlators of primary operators in a particular kinematic limit. The Regge regime of holographic four-point correlators was further studied from different points of view in~\cite{Cornalba:2007fs,Cornalba:2009ax,Costa:2012cb,Costa:2017twz,Kulaxizi:2017ixa,Li:2017lmh}.

A slightly different setup is to consider a fixed-target experiment in which a highly energetic particle scatters off a classical object whose mass is much larger than the energy of the incident test particle. A black hole is a prototypical example of such a heavy object. An interesting possibility, one that arises when considering a UV complete theory of gravity, is to consider a specific heavy {\it pure} state in place of the black hole. For instance, in the context of flat space type II string theories, the target can be represented by a stack of $N$ D$p$-branes~\cite{D'Appollonio:2010ae} and the Regge limit defined in a similar fashion to that of light $2\to 2$ scattering. A detailed comparison can then be made between the eikonal obtained from an amplitude approach and the dynamics of an energetic string probe propagating in the geometry produced by the D$p$-branes. In the AdS/CFT setup, the fixed-target version of the Regge limit was first studied in~\cite{Kulaxizi:2018dxo,Karlsson:2019qfi}, with the bulk heavy object represented by an asymptotically AdS$_{d+1}$ black hole -- or for $d=2$, a conical defect. On the CFT side, the heavy object is described by a state whose conformal dimension ($\Delta$) scales with the central charge $c$ of the CFT. The key observable in this case is a four-point CFT correlator involving two heavy and two light states (in the latter states, the $\Delta$ do not depend on $c$). This type of mixed heavy-light four-point correlator is usually dubbed HHLL. The analysis of~\cite{Kulaxizi:2018dxo,Karlsson:2019qfi} shows explicitly that in order to reproduce the result of the bulk calculation in the presence of a black hole, it is sufficient to characterise the heavy state by its couplings with the stress tensor and its multi-particle (``multi-trace'') versions.

The main aim of this paper is to apply the analysis of~\cite{Kulaxizi:2018dxo,Karlsson:2019qfi} to a heavy target that is an explicit pure state of large conformal dimension and to observe if and how the CFT data relevant for the eikonal depends on the choice of this pure state. Arguably the simplest setup that facilitates this aim is provided by the AdS$_3$/CFT$_2$ duality relevant for the D1-D5 CFT describing the prototypical example of a black hole in string theory -- the Strominger-Vafa black hole \cite{Strominger:1996sh}. The related bulk description is given by type IIB string theory compactified on AdS$_3 \times S^3 \times {\cal M}$, with ${\cal M}$ being either $T^4$ or $K_3$. The dual description is in terms of an ${\cal N}=(4,4)$ superconformal theory with $SU(2)_L \times SU(2)_R$ R-symmetry. In the original paper on the AdS/CFT conjecture~\cite{Maldacena:1997re}, this duality was derived from consideration of the decoupling limit of a stack of $n_1$ D1-branes and $n_5$ D5-branes, yielding the above mentioned SCFT with central charge $c=6N$. A long-standing effort to construct the gravitational duals of pure heavy states in this theory has led to the discovery of large classes of horizon-less ``microstate geometries'' having the same asymptotic structure as the black hole, but with different infrared behaviours encoding microscopic details of the states (see for instance \cite{Lunin:2001jy,Bena:2006kb,Kanitscheider:2007wq,Bena:2016ypk} and \cite{Warner:2019jll} for a recent review). Despite these families of solutions not covering the whole ensemble of the Strominger-Vafa black hole, they do provide an explicit semi-classical mechanism with which to replace the naive horizon with microscopic structure consistent with unitarity \cite{Mathur:2009hf,Almheiri:2012rt}. Here we will consider for heavy states, specific $1/2$-BPS chiral primary operators (CPO) and $1/4$-BPS operators. These states are atypical in the statistical ensemble of states of fixed conserved quantum numbers; however, the advantage is that a precise dual description in terms of asymptotically AdS$_3 \times S^3 \times {\cal M}$ microstate geometries is known~\cite{Kanitscheider:2007wq,Bena:2015bea,Bena:2016ypk,Bena:2017xbt}.

The heavy states ${\cal O}_H$ we consider are multi-particle operators composed of a large number $N_b$ of mutually BPS light operators ${\cal O}_L$. For reasons of simplicity, we take all constituents of the ${\cal O}_H$ to be identical -- that is ${\cal O}_H \sim {\cal O}_L^{N_b}$ (hence their very atypical nature). In order to have a heavy state, in the sense introduced above, it is necessary to keep the ratio $N_b/N$ finite when taking the large $c=6N$ limit. Even in this HHLL setup, $N_b/N$ is a free parameter and so following~\cite{Kulaxizi:2018dxo,Karlsson:2019qfi}, we can take a perturbative approach and extract the eikonal order by order in $N_b/N$. Such an approach is well-adapted for the calculation of the HHLL correlators\footnote{For finite values of $N_b/N$ the calculation of the correlators in the $1/4$-BPS states requires some approximation: a WKB approach was used in \cite{Bena:2019azk,Bena:2020yii}.} and several explicit examples are known~\cite{Galliani:2016cai,Galliani:2017jlg,Bombini:2017sge,Bombini:2019vnc,Tian:2019ash}. We study the OPE decomposition of such correlators in the channel describing the fusion of a heavy and light state, producing an intermediate excited heavy state -- this we call the ``cross channel''. Similarly to the case of standard LLLL correlators~\cite{Cornalba:2006xm}, the anomalous dimensions of these heavy excited states are directly related to the eikonal operator~\cite{Kulaxizi:2018dxo}. Likewise, the analytic bootstrap approach to the Regge regime can be adapted from the light~\cite{Li:2017lmh} to the heavy case, and a systematic perturbative approach in $N_b/N$ set up~\cite{Karlsson:2019qfi}.

This paper focuses solely on the first order in $N_b/N$, at which the eikonal in the HHLL regime is derived for atypical heavy states of the type mentioned above. Despite similarities between the conical defect and the effective 3D geometries describing the heavy pure states, the resulting eikonals are {\it different} already at this order. We show that results obtained from CFT correlators are in perfect agreement with the eikonal derived by studying geodesics in the dual microstate geometry -- the properties of geodesics in microstate geometries have been studied from various perspectives also in \cite{Tyukov:2017uig,Bianchi:2017sds,Bianchi:2018kzy,Bena:2018mpb,Bianchi:2020des,Bena:2020iyw}. By following~\cite{Li:2017lmh,Karlsson:2019qfi} we study the relevant bootstrap relation and show that it is satisfied by a different set of CFT data than in the conical defect case~\cite{Kulaxizi:2018dxo}. In both situations the ``direct channel'' -- in which the two light operators are fused together -- contains the contribution of the Virasoro block of the identity, but dressed by a different set of double-trace operators. In fact, generic conical defect geometries are not dual to pure states and it would be interesting to understand whether the CFT data extracted from them are fully consistent solutions of the bootstrap relation. As an aside, let us highlight that setting $N_b=1$ in the HHLL correlators described above, as done in~\cite{Giusto:2018ovt,Rastelli:2019gtj,Giusto:2019pxc}, reproduces the correlators of all light states in AdS$_3 \times S^3$~\cite{Rastelli:2019gtj,Giusto:2020neo}, despite the two regimes being not obviously connected. We analyse the Regge conformal bootstrap also in this regime, providing an explicit AdS$_3$/CFT$_2$ example of the analysis in~\cite{Kulaxizi:2017ixa,Li:2017lmh} and showing that the information obtained in the Regge regime can be used to fix some CFT data for spin-2 operators that was left undetermined in~\cite{Giusto:2018ovt}.

We conclude the introduction with an outline of the structure of this paper. In Section~\ref{sec:bg} we summarise the background material useful for deriving the eikonal in the geometric approach -- by calculating an appropriate geodesic -- and in the holographic CFT language, where one employs the four-point correlators. In Section~\ref{sec:Defect} we review and further analyse the result of~\cite{Kulaxizi:2018dxo} where the heavy states represent the conical defect AdS$_3$ geometries. In Section~\ref{sec:100Geometry} we consider a simple, yet non-trivial, class of $1/2$-BPS states. We also discuss in this explicit example how the Regge limit involving the HHLL correlator and the purely light case differ; showing why the conformal data obtained in the two cases are not the same. In Section~\ref{sec:101} we apply the same approach to a class of $1/4$-BPS states. A summary of our results and their possible extensions are outlined in the concluding Section~\ref{sec:conclusions}. The Appendices give details on the computations of integrals necessary in the CFT analysis of the HHLL and LLLL correlators in Sections~\ref{sec:100Geometry} and~\ref{sec:101}.

\section{Background material}
\label{sec:bg}

In this section we summarise basic material needed for the calculation of the eikonal phase in the context of the AdS/CFT duality. While the approach is general, we are particularly interested in the case relevant to the decoupling limit of a D1-D5 brane system, and so our equations will be specialised to the AdS$_3$/CFT$_2$ duality. We first provide a short discussion of the geodesic problem relevant to the semiclassical bulk calculation and then summarise the technology that can be used to derive the eikonal from CFT four-point correlators.

\subsection{The Regge limit in the AdS${}_3$ description} 
\label{sec:BulkPS}

In the gravitational picture, we will focus on 3D geometries that arise from the dimensional reduction of asymptotically AdS$_3 \times S^3$ solutions that are holographically dual to known CFT$_2$ heavy operators. We will need to consider the time delay and angular shift accrued by a null geodesic -- approximating the high energy light probe -- that begins and ends on the AdS boundary. As usual, by an appropriate choice of the affine parameter $\tau$, the equations for a null geodesic can be derived from the action\footnote{We use the mostly plus convention for the metric throughout.}
\begin{equation}
    S = \int\! d\tau\, \frac{1}{2}\, \frac{dx^{\mu}}{d\tau}\frac{dx^{\nu}}{d\tau}g_{\mu\nu} \ .
\end{equation}
We will focus on geometries with two Killing vectors; these are associated to the coordinates that, at the boundary, are identified with the temporal ($t$) and spatial ($y$) directions of the CFT. Thus, the momenta $p_{\mu}$ 
\begin{equation}\label{eq:conserved}
    p_{\mu} \equiv \frac{\delta S}{\delta \dot{x}^{\mu}} = g_{\mu\nu} \dot{x}^{\nu} \quad,\quad  \mbox{with } \quad\mu=t,y
\end{equation}
are conserved along the worldline (as usual a dot signifies a derivative with respect to $\tau$). From these, the phase shift can be defined as
\begin{align} \label{eq.BulkPS}
    \delta(\mathbf{p}) &\equiv - \mathbf{p}\cdot\Delta\mathbf{x}  =  -p_t \,\Delta t - p_y \,\Delta y \ , 
\end{align}
where $\Delta\mathbf{x}$ denotes the variation of the boundary coordinate $\mathbf{x}$ between the two ends of the geodesic. 
The geometries we consider can be written in a coordinate system where the metric is block-diagonal, i.e. the mixed components involving the radial direction and $t,y$ vanish. Then, the condition for a null geodesic $\dot{x}^\mu g_{\mu\nu} \dot{x}^\nu=0$ can be rewritten in terms of the conserved quantities as follows
\begin{align} \label{eq.NullEOM}
g_{rr}\dot{r}^2 = p_t^{\,2}\, \frac{g_{yy}-2\beta g_{ty} + \beta^2 g_{tt}}{g_{ty}^{\,2} - g_{tt}g_{yy}}\,,
\end{align}
where $\beta$ is related to the impact parameter of the geodesic
\begin{equation}\label{eq:betadef}
\beta\equiv \frac{p_y}{p_t}\,. 
\end{equation}
Other commonly used parameters are $s$ and $L$, defined by\footnote{Note that $p_t$ is negative for future-pointing geodesics.}
\begin{align} \label{eq.ptpyRelations}
    |p_t| = \frac{s}{R_y}\,\cosh L \quad,\ \quad p_y= \frac{s}{R_y}\,\sinh L \quad \Rightarrow \quad \tanh L= \beta\,,
\end{align}
where $R_y$ is the radius of the CFT spatial direction $y$.
The radial turning point $r_0$ is given by the  largest real solution of the equation $\dot{r} = 0$ and so is derived by setting Eq.~\eqref{eq.NullEOM} to zero. The time-delay $\Delta t$ and angular shift $\Delta y$ are then given by
\begin{align} \label{eq.delay}
    \Delta t = 2\int_{r_0}^{\infty}\! dr\, \frac{\dot{t}}{\dot{r}} \quad , \quad \Delta y = 2\int_{r_0}^{\infty}\! dr\, \frac{\dot{y}}{\dot{r}} \ .
\end{align}
By using~\eqref{eq.NullEOM} and the conserved quantities~\eqref{eq:conserved}, the eikonal can be written in terms of the following integral
\begin{align} \label{eq.BulkPS3}
      \delta = 2\,|p_t| \!\int_{r_0}^{\infty}\!dr\,\frac{\dot{t} + \beta\dot{y}}{\dot{r}} = 2\,|p_t|\!\int_{r_0}^{\infty}\!dr\, \sqrt{g_{rr}} \,\sqrt{\frac{g_{yy}-2\beta g_{ty} + \beta^2 g_{tt}}{g_{ty}^2 - g_{tt}g_{yy}}\,} \ .
\end{align}

\subsection{The Regge limit in the CFT${}_2$ description} \label{sec:CFT}

In the CFT picture, the eikonal is derived from four-point correlators containing two pairs of conjugate operators with dimensions $h_1$ and $h_2$
\begin{align} \label{eq.CorrUnfix}
    \langle \bar{\mathcal{O}}_1(z_1,\bar{z}_1){\mathcal{O}}_1(z_2,\bar{z}_2)\mathcal{O}_2(z_3,\bar{z}_3)\bar{\mathcal{O}}_2(z_4,\bar{z}_4)\rangle = z_{12}^{-2h_1}z_{34}^{-2h_2} \zb_{12}^{-2\hb_1}\zb_{34}^{-2\hb_2}\,G(z,\zb) \ ,
\end{align}
where $G$ is a function only of the conformal cross-ratios
\begin{equation} \label{eq.CrossRatios}
    z = \frac{z_{14}z_{23}}{z_{13}z_{24}} \ \ , \quad \zb = \frac{\zb_{14}\zb_{23}}{\zb_{13}\zb_{24}} \ ,
\end{equation}
and $z_{ij}\equiv z_i-z_j$. Upon fixing the positions of three operator insertions to $z_1=0$, $z_2=\infty$ and $z_3=1$ using conformal symmetry, this correlator can be written as
\begin{align} \label{eq.CorrExp}
    C(z,\zb) \equiv \langle {\mathcal{O}}_1| \mathcal{O}_2(1)\bar{\mathcal{O}}_2(z,\zb)|\bar{\mathcal{O}}_1\rangle &= (1-z)^{-2h_2}(1-\zb)^{-2\hb_2} \,G(z,\zb) \ ,
\end{align}
where $z_4=z$ and $\langle\mathcal{O}|\equiv\lim_{z\to\infty}z^{2h}\langle0|\mathcal{O}(z)$ is the BPZ conjugate state. The conformally invariant function $G(z,\zb)$ can be expanded in a basis of global conformal blocks, either in the direct channel ($z\to1$)
\begin{center}
\includegraphics[scale=.7]{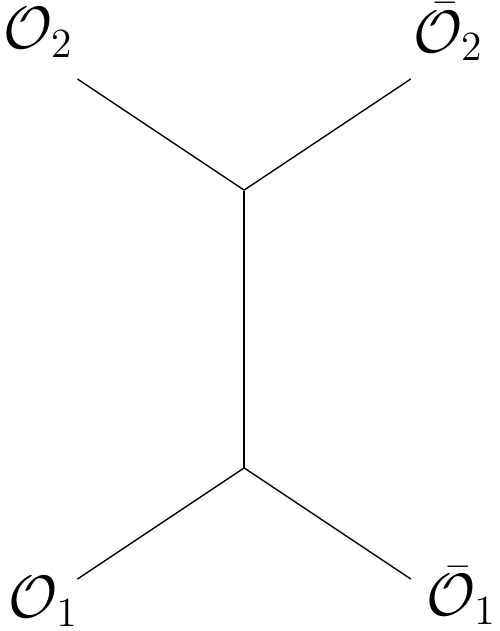}
\end{center}
or in the cross channel ($z\to0$)
\begin{center}
\includegraphics[scale=.75]{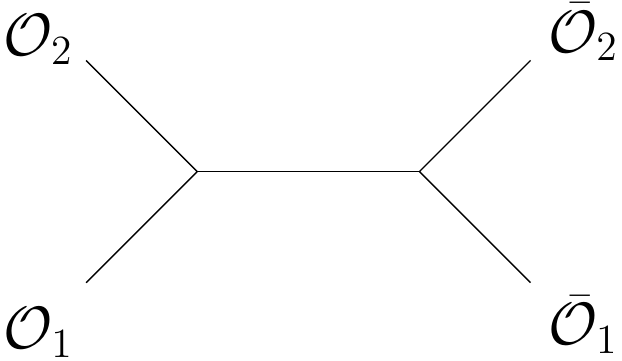}
\end{center}
These two different expansions of $ C(z,\zb)$ can be written as a sum over quasi-primary exchanges
\begin{equation} \label{eq.channelTS} 
     C(z,\zb) =\sum_{\mathcal{O}'} \frac{C_{11\mathcal{O}'}C_{\bar{\mathcal{O}}'22}}{(1-z)^{2h_2} (1-\bar{z})^{2\bar{h}_2}}\, g^{0,0}_{h,\hb}(1-z,1-\zb) = \sum_{\mathcal{O}} \frac{C_{12\mathcal{O}}C_{\bar{\mathcal{O}}21}}{z^{h_1+h_2} \bar{z}^{\bar{h}_1+\bar{h}_2}}\, g^{h_{12},\bar{h}_{12}}_{h,\hb}(z,\zb)\ ,
\end{equation}
where $h_{ij}\equiv h_i-h_j$ and the global conformal blocks, which resum the contributions of a quasi-primary with conformal dimensions $h$ and $\hb$ along with its infinite tower of global descendants, are given by
\begin{equation} \label{eq.ConformalBlockDef}
  g^{a,\bar a}_{h,\hb}(z,\zb) = z^{h}\zb^{\hb} {}_2F_1\big(h-a,h-a;2h;z\big)\,{}_2F_1\big(\hb-\bar a,\hb-\bar a;2\hb;\zb\big)\,.
\end{equation}
Notice that $z$ and $\zb$ are related by complex conjugation only in Euclidean space; despite this we will keep the same notation also when considering the analytic continuation to the Regge Minkowskian sheet.

We will be focusing on a holographic CFT$_2$ in the gravity regime, i.e. at large values of the central charge $c=6N$ and at strong coupling. The spectrum then contains a set of operators dual to the single-trace supergravity modes, while the stringy states decouple. These single-trace operators are light since their dimensions are of order one (as $c\to\infty$) and can be used to construct multi-trace operators. For instance, by using two single-particle operators ${\cal O}_i$ and ${\cal O}_j$, one can construct a family of quasi-primary double-trace operators that we write schematically as
\begin{align} \label{eq.DtraceOPs}
    \mathcal{O}_{ij} \equiv\, :\mathcal{O}_i\,\pd^m\pdb^{\mb}\mathcal{O}_j: \ .
\end{align}
These operators are labelled by the non-negative integers $m,\mb$ and have conformal dimensions of the form
\begin{equation} \label{eq.DTdims}
    h = h_i + h_j + m + \frac{1}{2}\gamma_{m,\mb}\;,\quad\quad
    \hb = \hb_i + \hb_j + \mb + \frac{1}{2} \gamma_{m,\mb} \ ,
\end{equation}
where $\gamma_{m,\mb}$ are the anomalous dimensions that are generically present when ${\cal O}_{ij}$ is not globally BPS -- even if the two single-particle constituents are individually protected. In the supergravity limit, the anomalous dimensions are suppressed in $1/N$ and so are small when compared to the leading contribution in~\eqref{eq.DTdims}: this is the starting point for the usual perturbative approach discussed below\footnote{An important detail here is that there is a degeneracy in the leading order spectrum which is (partially) lifted by the first order anomalous dimensions. We will not study how this lifting works and, with an abuse of notation, will use $\gamma$ to indicate the average anomalous dimension of a set of degenerate operators that appear in the OPE decomposition as discussed below.}. In the following we will often denote by $\ell = \abs{h-\hb}=\abs{m-\mb}$ the spin of the operator ${\cal O}_{ij}$, while the number of boxes ($\pd \pdb$) is given by $\min(m,\mb)$. 

We will need another class of multi-trace operators made from a large number $N_b\sim N$ of identical single-particle states, ${\cal O}_H \sim  {\cal O}_L^{N_b}$. These operators are ``heavy'', since their dimensions are of order $c$, and are dual to known asymptotically AdS$_3 \times S^3$ geometries. From the CFT point of view they behave as standard local operators; however, in order to highlight the effect of their large dimension in our correlators we will use upper case letters for the relevant quantum numbers. Thus, $H_i$ and $J_i$ will indicate the conformal weight and $U(1)\subset SU(2)_L$ R-charge (for notational simplicity we focus on the holomorphic part, but of course the discussion equally holds for the anti-holomorphic sector). In order to disentangle the Virasoro and $U(1)$ parts, it is convenient to introduce the ``reduced'' dimension of a heavy operator ${\cal O}_i\,$, in which the Sugawara $U(1)$ contribution is subtracted to give
\begin{equation}
H_i^{[0]}\equiv H_i - \frac{J_i^2}{N}\,.
\end{equation}
The class of heavy operators that we will consider has
\begin{equation}
H_i = N_b \left( n+ \frac{1}{2}\right) \,,\quad J_i = \frac{N_b}{2}\,,
\end{equation}
with $n$ a non-negative integer, and thus a reduced dimension of
\begin{equation} \label{eq.HeavyDimn}
H_i^{[0]}=N_b \left( n+ \frac{1}{2}- \frac{N_b}{4\,N}\right)\,.
\end{equation}
These heavy CFT states can be seen as part of an ensemble describing a black hole -- or more generally, a singular geometry such as a conical defect. The reduced conformal dimension is related to the mass $\mu$ of the underlying black hole by the relation
\begin{equation}\label{eq.alpha}
\sqrt{1-\mu\,} \equiv \alpha = \sqrt{1-\frac{24 \,H_1^{[0]}}{c}\,}= \sqrt{1-\frac{4\,N_b}{N}\left(n+\frac{1}{2}-\frac{N_b}{4\,N}\right)}\ ,
\end{equation}
where the central charge $c=6N$ was used. In this work we focus on the limit of small $N_b/N$ where one has
\begin{equation}\label{eq.mu}
\mu = 4\left(n+\frac{1}{2}\right) \frac{N_b}{N} + O\left ( \frac{N_b}{N}\right)^2\,.
\end{equation} 
To facilitate comparison with the literature, we will use $\mu$ as our expansion parameter in everything that follows.

Let us now go back to the analysis of the four-point correlator~\eqref{eq.CorrExp}: in the HHLL case we will take ${\cal O}_1$ to be the heavy state and ${\cal O}_2$ to be the light state, making the sets $\{\mathcal{O}\}$ and $\{\mathcal{O}'\}$ -- involved in the bootstrap relations~\eqref{eq.channelTS} -- qualitatively different. The dominant contribution in the direct channel is from the identity, on top of which there are single and double-trace light operators; whilst in the cross channel there will be no single-trace exchanges, but a tower of double-trace operators $\{\mathcal{O}_{12}\}$ -- again of the type in Eq.~\eqref{eq.DtraceOPs}, but involving a heavy and a light state. Heavy-light double-traces of this type, which we will also refer to by ${\cal O}_{\!H\!L}$, have dimensions 
\begin{equation} \label{eq.DTDims}
    H = H_1 + h_2 + m + \frac{1}{2}\Gamma_{m,\mb}\ \ \ ,\quad\ \
    \bar{H} = \bar{H}_1 + \hb_2 + \mb + \frac{1}{2} \Gamma_{m,\mb} \ .
\end{equation}
Here the $\Gamma_{m,\mb}$ are enhanced by a factor of $N_b$ with respect to the anomalous dimensions appearing in~\eqref{eq.DTdims}. Hence, the perturbative expansion of such heavy quantities will be in terms of $\mu\sim N_b/N$~\eqref{eq.mu}. Intuitively one can think of $\gamma_{m,\mb}$ as the binding energy between the two single particle constituents and $\Gamma_{m,\mb}$ accounting for the interaction of ${\cal O}_2$ with all constituents of the heavy operator ${\cal O}_1$. This picture holds only at first order in the ratio $N_b/N$, since in general the binding energies for the heavy/light bound states depend non-linearly on this ratio -- see for instance~\eqref{eq.BindingEnergy}. In this paper we will stick to this approximation and work at first order in $N_b/N$.

The strategy for analysing the HHLL correlators will be to expand the supergravity result at leading order in $\mu\sim N_b/N$ and to read off the CFT data relevant for Eq.~\eqref{eq.channelTS}. In this approximation we can use, for the operators entering in the cross channel, the expansions
\begin{align} \label{eq.muExpansions}
    \Gamma_{m,\mb} &= \mu\, \Gamma_{m,\mb}^{(1)} + \mu^2\, \Gamma^{(2)}_{m,\mb} +\cdots \nonumber\\
    C^{\,2}_{m,\mb} &\equiv C_{ij\mathcal{O}_{ij}} C_{\mathcal{O}_{ij}ij} = C^{\,2}_{\!(0)}(m,\mb)\,\bigg( 1+\mu\, C^{\,2}_{\!(1)}(m,\mb) + \mu^2\, C^{\,2}_{\!(2)}(m,\mb) +\cdots\bigg) \ .
\end{align}
At zeroth order in $\mu$, only the identity contributes to the direct channel and the bootstrap constraint~\eqref{eq.channelTS} reads
\begin{equation} \label{eq.N0crossing}
    \left.C(z,\zb)\right|_{\mu^0}= (1-z)^{-2h_2}  (1-\bar z)^{-2 \bar  h_2} = z^{-(H_1+h_2)} \bar{z}^{-(\bar{H}_1+\bar{h}_2)} \!\! \sum_{\{\mathcal{O}_{12}\}} \!C^{\,2}_{\!(0)}(m,\mb)\, g^{H_{12},\bar{H}_{12}}_{H,\bar{H}}(z,\zb)|_{\mu^0}  \ ,
\end{equation}
where $H_{12}=H_1-h_2$, $\bar{H}_{12}=\bar{H}_1-\bar{h}_2$. The generalised free field OPE coefficients $C^{\,2}_{\!(0)}$ are known~\cite{Heemskerk:2009pn}
\begin{align} \label{eq.OPEcoeff}
    C^{\,2}_{\!(0)}(m,\mb) = & \;\frac{\Gamma(2H_1+m) \Gamma(2h_2+m)  \Gamma(2H_1+2h_2 +m-1)}{m!\, \Gamma(2H_1) \Gamma(2h_2) \Gamma(2H_1+2h_2 +2 m-1)} \\ \nonumber & \times~ \frac{\Gamma(2\bar{H}_1+\bar{m}) \Gamma(2\bar{h}_2+\bar{m})  \Gamma(2\bar{H}_1+2\bar{h}_2 +\bar{m}-1)}{\bar{m}!\, \Gamma(2\bar{H}_1) \Gamma(2\bar{h}_2) \Gamma(2\bar{H}_1+2\bar{h}_2 +2 \bar{m}-1)}\ .
\end{align}
The same strategy can be used to analyse correlators in which all external operators are light~\cite{Li:2017lmh}, and in this case the expansion parameter is simply the inverse of the central charge, parametrised by $N^{-1}$. In our case the associated CFT data can then be expanded as
\begin{align} \label{eq.NExpansions}
    \gamma_{m,\mb} &= \frac{1}{N}\, \gamma_{m,\mb}^{(1)} + \frac{1}{N^2}\, \gamma^{(2)}_{m,\mb} +\cdots \nonumber\\
    c^{\,2}_{m,\mb} &\equiv c_{ij\mathcal{O}_{ij}} c_{\mathcal{O}_{ij}ij} = c^{\,2}_{(0)}(m,\mb)\,\left( 1+\frac{1}{N}\, c^{\,2}_{(1)}(m,\mb) + \frac{1}{N^2}\, c^{\,2}_{(2)}(m,\mb) +\cdots\right) \ ,
\end{align}
with $c^{\,2}_{(0)}(m,\mb)$ the leading order OPE coefficients.

Looking now to the cross channel decomposition in~\eqref{eq.channelTS} for the HHLL correlator at order $\mu$, we have
\begin{align} \label{eq.mu1Crossing}
    \left.C(z,\zb)\right|_{\mu} &= z^{-(H_1+h_2)}\, \bar{z}^{-(\bar{H}_1+\bar{h}_2)}\! \sum_{\{\mathcal{O}_{12}\}}\left. \!C^{\,2}_{m,\mb}\ g^{H_{12},\bar{H}_{12}}_{H,\bar{H}}(z,\zb)\right|_{\mu} \ .
\end{align}
On the right-hand side, the $\mu$ dependence is in both the OPE coefficients and the blocks (due to the anomalous dimensions). One difficulty in solving this constraint is that the first order corrections to both the OPE coefficients and the conformal dimensions appear as unknowns. In order to decouple their contributions and to make a connection to the classical bulk scattering of section \ref{sec:BulkPS}, we consider \eqref{eq.mu1Crossing} in the Regge limit. This limit involves analytically continuing around the origin one of the cross-ratios -- chosen to be $z$ -- to a second sheet, and then sending both $z$ and $\bar z$ to $1$:
\begin{align} \label{eq.ReggeLimit}
    z\to e^{-2\pi i}z \ \ \ \text{followed by }\ \ z,\zb\to1 \ .
\end{align}
It is helpful to parametrise the cross-ratios on the second sheet by $\sigma$ and $\eta$, with
\begin{align} \label{eq.zzbParam}
    z = 1-\sigma \ \ ,\quad \zb = 1- \sigma\,\eta \ ,
\end{align}
so that the Regge limit corresponds to sending $\sigma\to 0$ whilst keeping $\eta$ fixed. The order $\mu$ crossing equations \eqref{eq.mu1Crossing} in the Regge limit then read
\begin{align}\label{eq.mu1cc}
    \left.C_{\!\lcirclearrowright}\right|_{\mu} &= z^{-(H_1+h_2)}\, \bar{z}^{-(\bar{H}_1+\bar{h}_2)}\sum_{\{\mathcal{O}_{12}\}}\left. C^{\,2}_{m,\mb}\,e^{-2\pi i(H-H_1-h_2)} g^{H_{12},\bar{H}_{12}}_{H,\bar{H}}(z,\zb)\right|_{\mu} \\ \nonumber
    &= z^{-(H_1+h_2)}\, \bar{z}^{-(\bar{H}_1+\bar{h}_2)} \!\sum_{m,\mb=0}^{\infty} \left. C^{\,2}_{\!(0)}\bigg[ C^{\,2}_{\!(1)} +\frac{1}{2}\Gamma^{(1)}_{m,\mb}\Big(\!-2\pi i + (\pd_m+\pd_{\mb})\Big) \bigg] \,g^{H_{12},\bar{H}_{12}}_{H,\bar{H}}(z,\zb)\right|_{\mu=0}   \ ,
\end{align}
where the imaginary contribution follows from the factor of $z^H$ in the global blocks~\eqref{eq.ConformalBlockDef}. Selecting then the imaginary part of the above equation extracts a term proportional to the anomalous dimension and with no dependence on $C^{\,2}_{\!(1)}$:
\begin{equation} \label{eq.mu1Crossing2}
    \left. \text{Im}\;C_{\!\lcirclearrowright}\right|_{\mu} = -\pi z^{-(H_1+h_2)}\, \bar{z}^{-(\bar{H}_1+\bar{h}_2)}\! \sum_{m,\mb=0}^{\infty} C^{\,2}_{\!(0)}(m,\mb)\,\Gamma_{m,\mb}^{(1)}\left. g^{H_{12},\bar{H}_{12}}_{H,\bar{H}}(z,\zb)\right|_{\mu=0} \ .
  \end{equation}

The direct channel expansion~\eqref{eq.channelTS} includes the contribution of the ``universal'' sector consisting of: the identity, stress tensor and R-symmetry currents. These operators and their descendants contribute a universal part to the correlator, in the sense that it is completely determined by the symmetry algebra of the CFT -- depending only on the dimensions $H_1$, $h_2$, and $U(1)$ charges $J_1$, $j_2$ of ${\cal O}_1(\infty)$ and ${\cal O}_2(z)$. This universal contribution is given (for $N$ large and fixed $N_b/N$) by the product ${\cal V} = {\cal V}_V {\cal V}_A$ of the ``reduced'' Virasoro block of the identity \cite{Fitzpatrick:2015zha,Fitzpatrick:2014vua}
\begin{equation}\label{eq:virasoroblock}
\mathcal{V}_V(z) = z^{h_2 (\alpha-1)}\left( \frac{\alpha}{1-z^\alpha}\right)^{2 h_2}
\end{equation}
and the affine $U(1)$ block
\begin{equation}\label{eq:u1block}
\mathcal{V}_A(z)=  z^{\frac{2 J_1 j_2}{N}}
\end{equation}
(times the corresponding anti-holomorphic counterparts). On top of this ``universal'' sector, the direct channel contains a family of light double-trace operators $\{\mathcal{O}_{22}\}$ of the form~\eqref{eq.DtraceOPs} with $i=j=2$. The leading-order OPE coefficients $c^{\,2}_{\!(0)}$ of the double-trace operators in the direct channel are proportional to $N_b$, while the expansion parameter of the CFT data is $N^{-1}$ (see~\eqref{eq.NExpansions}), so at the first subleading order one reconstructs $N_b/N\sim \mu$ necessary to match the scaling of the cross channel~\eqref{eq.mu1cc}. The analysis in the direct channel is then essentially the same for the HHLL and the LLLL correlators: in both cases only single-trace or double-trace operators composed of two light constituents are exchanged, while all other multi-trace operators are suppressed in the large $N$ limit. In the LLLL case, it is then sufficient to simply set $N_b=1$. The perturbative expansion of the direct channel decomposition in~\eqref{eq.channelTS} then reads
\begin{align}
  \label{eq:finv}
  C(z,\bar{z}) = &\; {\cal V}(z) {\cal V}(\bar{z}) + \sum_{m,\bar{m}} c^{\,2}_{(0)}\!({m,\bar{m}})\, (1-z)^{m} (1-\bar{z})^{\bar{m}} F_m(z) F_{\bar m}(\bar{z})   \\ \nonumber + & \frac{\mu}{2} \sum_{m,\bar{m}}  (1-z)^{m} (1-\bar{z})^{\bar{m}} \Big[ \bar{\delta}(m,\bar{m}) \Big(\widehat{F}_m(z) F_{\bar m}(\bar{z}) + F_m(z)  \widehat{F}_{\bar m}(\bar{z})\Big) \\ \nonumber & \quad +\, \left(c^{\,2}_{(1)}({m,\bar{m}}) + \bar{\delta}(m,\bar{m}) \log\abs{1-z}^2 \right) F_m(z) F_{\bar m}(\bar{z}) \Big] + \dots\;,
\end{align}
where the $F$'s indicate the conformal block with $h=m+2 h_2\,$, $\bar{h}=\bar{m}+2 \bar{h}_2$ and its derivatives
\begin{equation}
  \label{eq:VdV}
  F_m(z) = {}_{2}F_1 (m+2h_2,m+2 h_2;2m+4h_2;1-z)\;,~~~
  \widehat{F}_m(z) = \partial_m F_m(z)\;.
\end{equation}
Since in the direct channel it is possible to have a vanishing average of the leading order OPE coefficients $c^{\,2}_{\!(0)}$, we have introduced the quantity $\bar{\delta} \equiv \langle c^{\,2}_{\!(0)}\gamma^{(1)}\rangle$ which is generically not equal to the product of the averages of $c^{\,2}_{\!(0)}$ and $\gamma^{(1)}$.

Due to the branch cut along $(\!-\infty,0]$ of the hypergeometric function ${}_2F_1(h,h\,;2h\,;1-z)$, present in the blocks $g^{0,0}_{h,\hb}$, the direct channel correlator will transform non-trivially upon moving to the second sheet relevant for the Regge limit. Using the analytic continuation across the branch cut yields  
\begin{align} \label{eq.2F1Monodromy}
    {}_2F_1(h,h;2h;1-z) \xrightarrow{\lcirclearrowright} {}_2F_1(h,h;2h;1-z) + 2\pi i\, \frac{\Gamma(2h)}{\Gamma^2(h)}\,{}_2F_1(h,h;1;z) \ .
\end{align}
Focusing on the imaginary part, the leading behaviour of a single direct channel global block in the $\sigma\to 0$ limit is then
\begin{align}\label{eq.GlobalRegge}
    \left. g^{0,0}_{h,\hb}(1-z,1-\zb)\right|_{\lcirclearrowright} \approx 2\pi i\, \frac{\Gamma\big(2 h \big)\,\Gamma\big(2 h -1\big)}{\Gamma^4(h)} \,\eta^{\bar{h}}\sigma^{1-h+\hb} \ ,
\end{align}
showing that operators with $h-\bar{h}$ large ({\it i.e.} large spin states) dominate. In particular, the spin-1 R-charge contribution, i.e. the $U(1)$ affine block in~\eqref{eq:u1block}, is subdominant with respect to the Virasoro block~\eqref{eq:virasoroblock}, which originates from the exchange of the stress-tensor. In our explicit examples we will see two different patterns. A first possibility is that operators with at most spin two are exchanged in the direct channel, such as the stress-tensor and the double-trace operators with $m=\bar{m}+2$. In this case, the analytic continuation to the Regge regime can be performed block by block, using~\eqref{eq.GlobalRegge} at leading order. Another possibility is to have contributions in the direct channel with unbounded spin: it is then necessary to first resum the terms with $m> \bar{m}+2$ and to perform the Regge analytic continuation on the result. We will later show how this is done in an explicit example (see section~\ref{ssec:light100} from \eqref{eq:G1111ef} onwards). In both cases, this direct channel analysis reproduces the Regge behaviour, {\it i.e.} the imaginary part of the correlator scales as $\sigma^{-2h_2 -1}$ in the $\sigma\to 0$ limit -- the extra factor of $-1$ in the exponent is typical of the exchange of a spin-2 state, identified holographically with the graviton.

By matching the $O(\mu)$ cross channel expansion on the r.h.s. of~\eqref{eq.mu1Crossing2} with the imaginary part of the correlator after having taken the Regge limit, one can extract the anomalous dimensions $\Gamma_{m,\mb}^{(1)}$ for operators with $m,\bar m\gg 1$ -- those dominating in the Regge regime. At this stage, a number of simplifying approximations can be made for both the OPE coefficients $C^{\,2}_{\!(0)}$ and the conformal blocks $g^{H_{12},\bar{H}_{12}}_{H,\bar{H}}(z,\zb)$; these approximations are different for the HHLL and LLLL cases.

HHLL correlators have $H_1\gg h_2,m, \bar m$: in this limit the OPE coefficients \eqref{eq.OPEcoeff} simplify to
\begin{equation} \label{eq.OPEcoeffH}
    C^{\,2}_{\!(0)}(m,\mb) \approx \frac{\Gamma(2 h_2+m)\,\Gamma(2\hb_2+\mb)}{m! \,\bar m!\,\Gamma(2h_2)\,\Gamma(2\hb_2)} \ ,
\end{equation}
and for $m,\bar m\gg 1$, relevant in the Regge limit, this further reduces to
\begin{equation} \label{eq.OPEcoeffHR}
    C^{\,2}_{\!(0)}(m,\mb) \approx  \frac{m^{2 h_2-1} \mb^{2 \hb_2-1}}{\Gamma(2h_2)\,\Gamma(2\hb_2)}  \ .
\end{equation}
The hypergeometric functions in the cross channel conformal blocks \eqref{eq.ConformalBlockDef} can also be approximated in the limit $H_1\gg h_2,m, \bar m$ by
\begin{equation} \label{eq.HypergeoApprox}
    {}_2F_1(H-H_{12},H-H_{12};2H;z) = \sum_{k=0}^{\infty} \frac{\poc{H-H_{12}}{k}^{\,2}}{k!\poc{2 H}{k}}\,z^k \approx 1 + O(1/H_1) \ ,
\end{equation}
where we used the series representation of the hypergeometric function and the approximations $H\approx H_1$, which follows from \eqref{eq.DTDims}, and $H_{12}=H_1-h_2\approx H_1$. Implementing these approximations in the Regge crossing equation \eqref{eq.mu1Crossing2} gives
\begin{equation} \label{eq.mu1CrossingH}
    \left. \text{Im}\;C_{\!\lcirclearrowright}\right|_{\mu^1} \approx -\pi \sum_{m,\bar m=0}^{\infty} C^{\,2}_{\!(0)}(m,\mb)\,\Gamma_{m,\mb}^{(1)} \,z^{m}\zb^{\mb} \ .
\end{equation}
 
For LLLL correlators, the cross channel decomposition in the Regge limit is identical to \eqref{eq.mu1Crossing2} with $C^{\,2}_{\!(0)}\to c^{\,2}_{\!(0)}$, $\Gamma_{m,\mb}^{(1)}\to \gamma_{m,\mb}^{(1)}$, $g^{H_{12},\bar{H}_{12}}_{H,\bar{H}}\to g^{h_{12},\bar{h}_{12}}_{h,\bar{h}}$ and $H_1\to h_1$, where the conformal dimension $h_1$ is of order 1 in the large $c$ limit. In this regime, the Regge limit allows for an approximation to the conformal blocks in terms of modified Bessel functions of the second kind -- since again, double-trace operators with large $m,\mb$ dominate in the cross channel. Thus, considering $h,\hb\gg1$ with $\hat{z}\equiv h\sqrt{1-z\,}$ finite, the holomorphic part of the conformal blocks \eqref{eq.ConformalBlockDef} approximates to \cite{Li:2017lmh}
\begin{align} \label{eq.BlockApproxL}
    z^{h} {}_2F_1\big(h-h_{12},h-h_{12};2h;z\big) \approx 2^{2h}\sqrt{\frac{h}{\pi}\,}\,(1-z)^{h_{12}} \,K_{-2h_{12}}\big(2h\sqrt{1-z\,}\big)\equiv {\cal K}^{\,h_{12}}_h \ ,
\end{align}
giving the full conformal block as
\begin{align}
    g^{h_{12},\bar{h}_{12}}_{h,\bar{h}} \approx \mathcal{K}^{\,h_{12}}_h(z) \mathcal{K}^{\,\bar h_{12}}_{\hb}(\zb)  \ .
\end{align}
We recall that the HHLL correlator at first order in $N_b/N$ and the LLLL correlator (which has $N_b=1$) at first order in $1/N$ are identical. Despite this fact, the approximations to the conformal blocks and the OPE coefficients that are appropriate in the two regimes are different: for the conformal blocks one should use \eqref{eq.HypergeoApprox} in the HHLL regime and \eqref{eq.BlockApproxL} in the LLLL one. For this reason the anomalous dimensions $\Gamma_{m,\mb}^{(1)}$ and  $\gamma_{m,\mb}^{(1)}$ that one derives in the two cases are different. This fact will be illustrated in a specific example in section \ref{ssec:light100}. 

Both anomalous dimensions $\Gamma_{m,\mb}^{(1)}$ and  $\gamma_{m,\mb}^{(1)}$ for large values of $m, \mb$ are linked to the $O(\mu)$ phase shift $\delta^{(1)}$ computed on the gravity side by identical relations \cite{Cornalba:2006xm,Kulaxizi:2017ixa}
\begin{equation} \label{eq.AnomDimPSmu}
  \Gamma_{m,\mb}^{(1)}\approx  -\frac{\delta^{(1)}}{\pi} \ \ ,\quad  \gamma_{m,\mb}^{(1)} \approx -\frac{\delta^{(1)}}{\pi} \quad \mathrm{for}\ \  m,\mb\gg 1\ ,
\end{equation}
where the CFT variables $m,\mb$ are mapped to the momenta $p_t, p_y$ -- of which $\delta^{(1)}$ is a function -- by
\begin{equation}\label{eq.sLRelations}
R_y\,|p_t| =m+\mb\ \ ,\quad R_y\,p_y =m-\mb  \quad \Rightarrow \quad \beta= - \frac{m-\mb}{m+\mb}\ .
\end{equation}

\section{The example of the conical defect geometry} \label{sec:Defect}

In order to make a connection with the AdS${}_3$ conical defect geometry analysed in \cite{Kulaxizi:2018dxo}, we consider a particularly simple microstate geometry, first introduced in \cite{Balasubramanian:2000rt,Maldacena:2000dr}. This 6D geometry locally factorises into AdS${}_3\,\times\,$S${}^3$ and for our purposes only the reduced 3D metric is relevant, given by
\begin{align}
 (Q_1Q_5)^{-\frac{1}{2}} ds^2_{\mathrm{AdS}_3} = \frac{dr^2}{r^2 + \tfrac{a^2}{k^2}\,} - \frac{r^2 + \tfrac{a^2}{k^2}\,}{Q_1Q_5} dt^2 + \frac{r^2}{Q_1Q_5} dy^2 \,,\label{eq.DefectGeoAdS3}
\end{align}
where $k\in\mathbb{N}$. The radius $R_y$ of the $y$ coordinate is related to the D1, D5 charges $Q_1$, $Q_5$ and the parameter $a$ by $R_y = \frac{\sqrt{Q_1 Q_5}}{a}$. With the periodic identification $y\sim y + 2\pi R_y\,$, the above geometry has a conical singularity of order $k$ at $r=0$. One could formally eliminate the conical singularity and map the metric \eqref{eq.DefectGeoAdS3} to global AdS${}_3$ by the local diffeomorphism $r\to r\,k^{-1}$, $t\to t\,k$, $y\to y\,k$. However, since this diffeomorphism is non-vanishing at the AdS boundary and is not globally defined due to the change in $y$ periodicity it induces, the geometry \eqref{eq.DefectGeoAdS3} and global AdS${}_3$ are physically inequivalent. 

The conical singularity has a natural description at the orbifold point of the dual D1-D5 CFT: the heavy operator dual to the geometry \eqref{eq.DefectGeoAdS3} is made up of $N/k$ copies of the twist operator of order $k$ \cite{Lunin:2001jy}. This description makes it evident that only the geometries with integer $k$ can be associated to states of the CFT. Nevertheless, in order to connect with \cite{Kulaxizi:2018dxo}, in which geometries with real-valued defect angles were considered, one can analytically continue $k$ to take generic values in $[1,\infty)$ and parametrise it as
\begin{align} \label{eq.kContinued}
    \frac{1}{k} = \sqrt{1-\mu\,} \equiv \alpha \ ,
\end{align}
where $\mu=0$ describes pure AdS. The bulk phase shift computed in the reduced 3D metric \eqref{eq.DefectGeoAdS3} follows from the general formula \eqref{eq.BulkPS3}:
\begin{align} \label{eq.DefectBulkPS}
    \delta_{k} = 2aR_y\,|p_t|\!\int_{r_0}^{\infty}\!dr\,\bigg(r^2+\frac{a^2}{k^2}\bigg)^{-1}\sqrt{1-\frac{\beta^2}{r^2}\bigg(r^2+\frac{a^2}{k^2}\bigg)} = \pi R_y\,k\,|p_t|(1-|\beta|) \ ,
\end{align}
with the radial turning point $r_0=\tfrac{a}{k}(\beta^{-2}-1)^{-1/2}$ obtained by setting Eq.~\eqref{eq.NullEOM} to zero. It is noted that setting $k\to1$ here reproduces the phase shift in pure AdS${}_3$ as expected. Subtracting the AdS result from \eqref{eq.DefectBulkPS} gives the deviation due to the presence of the defect as
\begin{align} \label{eq.DefectBulkPS2}
    \delta = \delta_k-\delta_{k=1}= \pi R_y\,|p_t|(1-|\beta|)\,(k-1) \ .
\end{align}
Using the analytic continuation \eqref{eq.kContinued}, the phase shift can be expanded in small $\mu$ allowing for a CFT interpretation of the bulk result and comparison with \cite{Kulaxizi:2018dxo}:
\begin{align} \label{eq.DefectBulkPSmu}
    \delta =\pi R_y\,|p_t|(1-|\beta|)\Bigl[(1-\mu)^{-\frac{1}{2}}-1 \Bigr] =\pi R_y\,|p_t|(1-|\beta|)\bigg( \frac{1}{2}\mu +\frac{3}{8}\mu^2 + \cdots\bigg) \ .
\end{align}

We would like to understand if the bulk phase shift \eqref{eq.DefectBulkPSmu} captures the Regge limit of some CFT correlator. For integer $k$ this would be the four-point correlator between the heavy state dual to the conical defect \eqref{eq.DefectGeoAdS3} and two light operators of fixed conformal dimension $(h_L,\bar h_L)$. This four-point correlator has been computed in \cite{Galliani:2016cai} by solving the linearised wave equation describing small fluctuations of the light operator in the background \eqref{eq.DefectGeoAdS3} of the heavy operator. When the light operator is taken to be the chiral primary operator $\mathcal{O}^{\mathrm{fer}}$ of dimension $(h_L,\hb_L)=(1/2,1/2)$, the correlator in the NSNS sector is
\begin{equation} \label{eq.DefectCorrfer}
    C^{\mathrm{fer}}_k= \frac{1/k}{|1-z|^2}\frac{1-|z|^2}{1-|z|^{2/k}} \ .
\end{equation}
Another natural candidate for the light operator is $\mathcal{O}^{\mathrm{bos}}$, with dimension $(h_L,\hb_L)=(1,1)$. This super-descendant of $\mathcal{O}^{\mathrm{fer}}$ is obtained by acting on the chiral primary with one left-moving and one right-moving supercharge. In the bulk, $\mathcal{O}^{\mathrm{bos}}$ has a simpler description than $\mathcal{O}^{\mathrm{fer}}$, being dual to a minimally coupled scalar in the background described by the 6D Einstein metric. The correlators $C^{\mathrm{fer}}$ and $C^{\mathrm{bos}}$ of the light operators $\mathcal{O}^{\mathrm{fer}}$ and $\mathcal{O}^{\mathrm{bos}}$ in a 1/2-BPS heavy state (such as the one dual to \eqref{eq.DefectGeoAdS3}) are related by a simple supersymmetric Ward identity, which gives
\begin{align} \label{eq.DefectCorrbos}
    C^{\mathrm{bos}}_k = \pd\pdb\,\Big[C^{\mathrm{fer}}_k\Big]=\pd\pdb \left(\frac{1/k}{|1-z|^2}\frac{1-|z|^2}{1-|z|^{2/k}}\right) \ .
\end{align}
To compare with the bulk phase shift computed in a conical defect geometry with real-valued deficit angle, one can analytically continue the above correlators using the parametrisation \eqref{eq.kContinued} to get
\begin{equation} \label{eq.DefectCorrferbos}
    C^{\mathrm{fer}}_\alpha= \frac{\alpha}{|1-z|^2}\frac{1-|z|^2}{1-|z|^{2\alpha}}\,,\quad  C^{\mathrm{bos}}_\alpha= \pd\pdb \left(\frac{\alpha}{|1-z|^2}\frac{1-|z|^2}{1-|z|^{2\alpha}}\right) \ .
\end{equation}
After analytic continuation, $C^{\mathrm{fer}}_\alpha$ and $C^{\mathrm{bos}}_\alpha$ can no longer be interpreted as correlators of a pure heavy state of the CFT. One possibility is that they represent correlators in an ensemble of 1/2-BPS states with an average conformal dimension set by the parameter $\alpha$~\eqref{eq.alpha}.
This identification is consistent with the lightcone OPE limit $\zb\to 1$ of the correlators. As an example, $C^{\mathrm{bos}}$ in this limit is given by \cite{Bombini:2017sge}
\begin{equation}
    C^{\mathrm{bos}}_\alpha\xrightarrow{\zb\to 1} \frac{z^{\alpha-1}}{(1-\zb)^2}\left(\frac{\alpha}{1-z^\alpha}\right)^2 \ ,
\end{equation}
which by comparison with \eqref{eq:virasoroblock}, is the HHLL Virasoro identity block with light operators of dimension $h_L=1$ (multiplied by the prefactor from \eqref{eq.CorrExp}). 

We now study the Regge limit of this correlator and, to help the  CFT interpretation, we also take the small $\mu$ expansion. Focusing on the first order in $\mu$, the imaginary part of the Regge limit of $C^{\mathrm{bos}}$ obtained after performing the analytic continuation~\eqref{eq.ReggeLimit} reads
\begin{equation}\label{eq.cdrl}
    \mathrm{Im} \,C^{\mathrm{bos}}_{_\alpha\,\circlearrowright}\Bigl|_{\mu^1} \approx \frac{2\pi}{\sigma^4 \eta^2} \left(\frac{1+3\eta+\eta^2}{\sigma (1+\eta)^3}\right) =  \frac{2\pi}{\sigma^4 \eta^2} \left(\frac{1-2 \eta + 5 \eta^2 - 9 \eta^4 + \ldots}{\sigma}\right) \ ,
  \end{equation}
where we used the parametrisation in \eqref{eq.zzbParam} and kept only the leading term in $\sigma$. The overall factor of $\sigma^{-4} \eta^{-2}$ comes from the prefactor $(1-z)^{-2 h_2} (1-\bar{z})^{-2 \hb_2}=\sigma^{-2 (h_2+\hb_2)} \eta^{-2\hb_2}$ in~\eqref{eq.CorrExp} with $h_2=\hb_2=1$. By expanding the remaining part of the result in small $\eta$ (as done in the second equality of \eqref{eq.cdrl}) one can gain some insight on the CFT meaning of the correlator $C^{\mathrm{bos}}_\alpha$. Comparing each term of the small $\eta$ expansion with the behaviour of the blocks in the Regge limit~\eqref{eq.GlobalRegge}, it is natural to interpret a contribution scaling like $\sigma^{-1} \eta^{n}$ for $n\geq0$ as being due to the exchange of primaries of weight $(h,\hb)=(2+n,n)$. In particular, taking the Regge limit of the Virasoro block of the identity produces only the first term in the small $\eta$ expansion.  As a consistency check of this interpretation, we can compare the first few coefficients of the $\eta$ expansion in \eqref{eq.cdrl} with those obtained in the Euclidean OPE decomposition as $z\to 1$ (given by~\eqref{eq:finv} before the analytic continuation needed for the Regge limit). From the first few terms in the Euclidean decomposition one can see the following pattern emerging: both the leading order couplings $c^{\,2}_{(0)}$ and the anomalous dimensions $\bar{\delta}$ are trivial, while for the couplings at order $\mu$ there are no contributions of spin higher than two. For instance, one can easily obtain the following data
\begin{equation}
  \label{eq.EopeCD}
\begin{aligned}
& c^{\,2}_{(1)}({0,0})=\frac{1}{30}\;,\quad c^{\,2}_{(1)}({1,1})=-\frac{1}{210}\;,\quad c^{\,2}_{(1)}({2,2})=\frac{1}{275} \;,\;\ldots\\
&  c^{\,2}_{(1)}({2,0})=-\frac{1}{700}\;,\quad c^{\,2}_{(1)}({3,1})=\frac{1}{4410}\;,\quad c^{\,2}_{(1)}({4,2})=-\frac{1}{38808}\;,\;\ldots\;\;,
\end{aligned}
\end{equation}
and of course $c^{\,2}_{(1)}({m,m+2})=c^{\,2}_{(1)}({m+2,m})$. The couplings of the states with spin $2$ agree with the expansion of the round parenthesis in~\eqref{eq.cdrl} once the normalisation in~\eqref{eq.GlobalRegge} is taken into account. This can be checked by multiplying the results in~\eqref{eq.EopeCD} by the factor present in~\eqref{eq.GlobalRegge}: for $m=2,3,4\ldots$
\begin{equation}
  \label{eq:313}
  \frac{\Gamma(2m+4)\,\Gamma(2m+3)}{\Gamma^4(m+2)}\,\frac{\mu}{2} \,c^{\,2}_{(1)}({m,m-2}) \to \mu\,(-2,5,-9,\ldots)\;.
\end{equation}

We now analyse the cross channel interpretation of~\eqref{eq.cdrl} using~\eqref{eq.mu1CrossingH}, which is dominated by the double-trace operators of the form $\mathcal{O}_{\!H} \partial^m {\bar \partial}^{\mb}\mathcal{O}_{\!L}$, with large values of $m$ and $\mb$. The anomalous dimensions $\Gamma^{(1)}_{m\mb}$  are encoded in the phase shift \eqref{eq.DefectBulkPSmu}, computed from the analytically continued conical defect geometry. From \eqref{eq.AnomDimPSmu} and the identifications \eqref{eq.sLRelations}, one finds that
\begin{equation} \label{eq.defectgamma1}
    \Gamma^{(1)}_{m,\mb} \approx -\min(m,\mb)\,,
\end{equation}
in agreement with \cite{Kulaxizi:2018dxo} (see also Eq.~(6.4) of~\cite{Li:2020dqm} which captures the large $h_H,\; \bar{h}_H$ limit of Eq.~(4.32) of~\cite{Kraus:2018zrn}). We can then resum the contributions of these double-trace operators with~\eqref{eq.mu1CrossingH} by approximating the sums with integrals and using~\eqref{eq.OPEcoeffHR} with $h_2=\hb_2=1$
\begin{equation}
  \label{eq:mcd}
  \begin{aligned}
    \mathrm{Im} \,C^{\mathrm{bos}}_{_\alpha\,\circlearrowright}\Bigl|_{\mu^1} = &\; \pi \left[\int_0^\infty \!\! dm \int_0^m \!\! d\mb \,  m\, \mb^2 \, z^m \bar{z}^{\mb} + \int_0^\infty \!\! d\mb \int_0^{\mb} \!\! dm \,  m^2 \mb \, z^m \bar{z}^{\mb}\right]
    \\  = &\; \pi \Big(I_{1,2,0}(z,\bar{z}) +  I_{1,2,0}(\bar{z},z) \Big) = 2\pi\, \left(\frac{1+3\eta+\eta^2}{\sigma^5 \eta^2 (1+\eta)^3}\right)\;,
\end{aligned}
\end{equation}
where in the second line we used the result~\eqref{eq.GeneralAsymInt} and reproduced the Regge behaviour \eqref{eq.cdrl}, including all terms of order $\sigma^{-1} \eta^{n}$ for $n\geq0$. Thus, while the Virasoro block of the identity alone does not provide a consistent solution to the bootstrap problem, the ``correlator'' $C^{\mathrm{bos}}_{\alpha}$ does. The terms $\sigma^{-1} \eta^{n}$ with $n\geq0$ originate from the double-trace primaries $\bar{\mathcal{O}}_L \partial^{2+n} {\pdb}^{n}\mathcal{O}_L$ exchanged in the direct channel ($z,\bar z\to 1$).

The same analysis can be performed for the analytically-continued correlator with light operator $\mathcal{O}^{\mathrm{fer}}$ given in \eqref{eq.DefectCorrferbos}. After the analytic continuation to the Regge region and the small $\mu$ expansion, the order $\mu$ contribution is
\begin{equation} \label{eq.DefectferRegge}
    \mathrm{Im} \,C^{\mathrm{fer}}_{\alpha\,\circlearrowright}\Bigl|_{\mu^1} \approx \frac{\pi}{\sigma^3 \eta(1+\eta)} \ .
\end{equation}
Of course, one can relate~\eqref{eq.DefectferRegge} and~\eqref{eq.cdrl} directly by writing the Ward identity~\eqref{eq.DefectCorrbos} in the variables $(\sigma,\eta)$ adapted to the Regge limit
\begin{equation}
    \label{eq:zbse}
    \pd = -\partial_\sigma+ \frac{\eta}{\sigma} \partial_\eta\ ,\ \ \pdb= -\frac{1}{\sigma}\partial_\eta \quad \Rightarrow \quad \left(\partial_\sigma- \frac{\eta}{\sigma} \partial_\eta\right) \left(\frac{1}{\sigma}\partial_\eta\right) \mathrm{Im} \,C^{\mathrm{fer}}_{\alpha\,\circlearrowright}\Bigl|_{\mu^1} = \mathrm{Im} \,C^{\mathrm{bos}}_{\alpha\,\circlearrowright}\Bigl|_{\mu^1} \ .
\end{equation}
For large values of $m$ and $\mb$, the anomalous dimensions $\Gamma^{(1)}_{m,\mb}$ of the double-trace operators contributing to the cross channel of $C^{\mathrm{fer}}_{\alpha}$ are equal to the ones extracted from $C^{\mathrm{bos}}_{\alpha}$. This agrees with the idea that the two light operators $\mathcal{O}^{\mathrm{fer}}$ and $\mathcal{O}^{\mathrm{bos}}$ are indistinguishable in the Regge limit, both being represented by null geodesics in the 3D spacetime. The couplings $C_{\!(0)}^{\,2}$ change simply due to the dimension of the light external operator now being $h_2=1/2$: using this value in~\eqref{eq.OPEcoeffH}, one obtains from the cross channel decomposition an integral with the same structure as in~\eqref{eq:mcd} but involving $I_{0,1,0}$ instead of $I_{1,2,0}$, which reproduces~\eqref{eq.DefectferRegge}.

Let us conclude this section with some comments. The analysis of~\cite{Kulaxizi:2018dxo} starts from the HHLL Virasoro vacuum block in the direct channel, then from this input the CFT data in the cross channel are derived. The contributions from double-trace operators $\mathcal{O}_{\!L} \partial^m {\bar \partial}^{\mb}\mathcal{O}_{\!L}$ in the direct channel are added as a final step in order to satisfy crossing. Here we start from $C_\alpha^{\mathrm{fer}}$ or $C_\alpha^{\mathrm{bos}}$ which already contain the exchanges of the operators $\mathcal{O}_{\!L} \partial^m {\bar \partial}^{\mb}\mathcal{O}_{\!L}$ and provide directly a solution to the crossing constraint as discussed in this section. In spite of this difference in starting point, the results for the anomalous dimensions in the Regge limit -- given in Eq.~\eqref{eq.defectgamma1} -- of the double-trace operators $\mathcal{O}_{\!H} \partial^m {\bar \partial}^{\mb}\mathcal{O}_{\!L}$ agree, implying that we are finding the same solution to the crossing constraint as in~\cite{Kulaxizi:2018dxo}. We emphasise that although $C^{\mathrm{bos}}_{\alpha}$ and $C^{\mathrm{fer}}_{\alpha}$ satisfy the bootstrap relation, we know from the argument given at the beginning of this section that they cannot represent correlators in pure states for generic real values of $\alpha$. This argument is based on the observation that the conical defect geometry \eqref{eq.DefectGeoAdS3} has an allowed conical singularity only for integer $k$. It would be interesting to understand if there are consistency requirements, detectable purely within the CFT, that are violated by $C^{\mathrm{bos}}_{\alpha}$ and $C^{\mathrm{fer}}_{\alpha}$ for generic values of $\alpha$.

\section{A class of two-charge microstate geometries} \label{sec:100Geometry}

We now consider the phase shift in the context of the D1-D5 system. Firstly, we focus on the simplest subset of heavy states; the 1/2-BPS heavy operators that are in correspondence (via spectral flow of the CFT) with the Ramond-Ramond ground states of the theory. Though the ensemble of these states does not give rise to a classical black hole with finite horizon, it still represents a non-trivial ensemble with a macroscopically large entropy. The simplest states in this ensemble are the duals of the conical defect geometries with integer $k$, given in \eqref{eq.DefectGeoAdS3}. On the CFT side those states are highly symmetric, being formed from many identical copies of one elementary constituent (a twist operator of the orbifold CFT) and this is reflected on the gravity side by the fact that the geometries are locally isomorphic to AdS${}_3\times S^3$. It is interesting to extend the analysis to more generic states that still allow for an analytic treatment. For instance, the $(k,0,0)$ family of solutions has tended to be a useful playground; these were first constructed in \cite{Kanitscheider:2007wq} and later provided the seed for the construction of \cite{Bena:2016ypk}.

\subsection{The bulk description}
\label{ssec:bulk}

The $(k,0,0)$ spacetimes cannot be factorised, even locally, into asymptotically AdS${}_3$ and asymptotically $S^3$ parts and thus have to be described in 6D. The full geometry is given, for example, in Eq.~(3.11) of \cite{Bena:2015bea}. It is useful, for our purposes at least, to rewrite the 6D Einstein metric in a ``dimensionally reduced" form
 \begin{equation} \label{eq.kkansatz}
    ds^{\,2}_6 = V^{-2}g_{\mu\nu}dx^{\mu}dx^{\nu} + G_{\alpha\beta}\big(dx^{\alpha} + A^{\alpha}_{\mu}dx^{\mu} \big) \big( dx^{\beta} + A^{\beta}_{\nu} dx^{\nu} \big) \ ,
\end{equation}
where $x^\mu, x^\nu$ denote the AdS${}_3$ coordinates $(r,t,y)$; $x^\alpha, x^\beta$ the $S^3$ coordinates $(\theta,\phi,\psi)$; $A^{\alpha}_{\mu}$ are $SO(4)$ gauge fields; and the metrics $g_{\mu\nu}$ and $G_{\alpha\beta}$ reduce at large $r$ to those of AdS${}_3$ and $S^3$ respectively. In \eqref{eq.kkansatz}, $V$ is a warping factor chosen in such a way to ensure that, when not dependant on the $S^3$ coordinates, $g_{\mu\nu}$ is the Einstein metric in 3D:
 \begin{equation}
    V^2 \equiv \frac{\det G_{\alpha\beta}}{\det G^{(0)}_{\alpha\beta}}\ ,
\end{equation}
with $G^{(0)}_{\alpha\beta}$ being the large $r$ limit of $G_{\alpha\beta}$ which, as mentioned, is the round unit $S^3$ metric (multiplied by $(Q_1 Q_5)^{1/2}$). While a reduction of the form \eqref{eq.kkansatz} can always be written down, in general the 3D reduced metric $g_{\mu\nu}$ will depend on both the $x^\mu$ and $x^\alpha$ coordinates at finite $r$. A simplification occurs for $k=1$; in this case $g_{\mu\nu}$ turns out to be $x^\alpha$ independent and thus can be thought of the Einstein metric of a 3D spacetime that is asymptotically, but not locally, AdS${}_3$. For $k=1$ one can thus reduce the 6D problem to a simpler 3D one and in the following we will restrict to the $(1,0,0)$ state to take advantage of this simplification.

Before giving the full form of the $(1,0,0)$ geometry, we clarify the set of parameters on which it depends: these are the D1, D5 charges $Q_1$, $Q_5$; the radius of the CFT circle $R_y$; and two parameters $a$ and $b$ constrained by the relation
\begin{equation}\label{eq:a0def}
    a^2 + \frac{b^2}{2} = \frac{Q_1 Q_5}{R_y^2} \equiv a_0^2\ .
\end{equation}
Therefore, the parameter $b$ can be varied whilst keeping the CFT quantities $Q_1$, $Q_5$ and $R_y$ fixed. In this way we get a continuous family of heavy states, all of which are collectively described by the $(1,0,0)$ solution. Specifically, $b$ is related to the number $N_b$ of single-particle constituents of the heavy state that are not the NSNS vacuum via
\begin{align}\label{eq:bN100}
    \frac{N_b}{N} = \frac{b^2}{2a_0^2} \ .
\end{align}
In particular, when $b=0$ we have $N_b=0$ and the state is just the NSNS vacuum, whose dual geometry is global AdS${}_3\times S^3$.

The explicit form of the $(1,0,0)$ solution is given by the asymptotically $S^3$ metric
\begin{equation}
    G_{\theta\theta}=\sqrt{\mathcal{P}\,} \,\Sigma\,,\, G_{\phi\phi}=\frac{Q_1 Q_5}{\sqrt{\mathcal{P}\,} \,\Sigma}\,\sin^2\theta\,,\,\,G_{\psi\psi}=\frac{Q_1 Q_5}{\sqrt{\mathcal{P}\,}\, \Sigma}\,\frac{r^2+\frac{a^4}{a_0^2}}{r^2+a^2}\,\cos^2\theta\,;
\end{equation}
the gauge fields
\begin{equation} \label{eq.3dA100}
    A^\theta=0\,,\,\, A^\phi = -\frac{a^2}{a_0^2} \frac{dt}{R_y}\,,\,\, A^\psi = -\frac{a^2}{a_0^2} \,\frac{r^2+a^2}{r^2+\frac{a^4}{a_0^2}} \frac{dy}{R_y}\,;
\end{equation}
and the 3D Einstein metric
\begin{align} \label{eq.3d100Metric}
    ds_3^{\,2} = g_{\mu\nu} dx^\mu dx^\nu = \sqrt{Q_1Q_5}\,\frac{r^2+\frac{a^4}{a_0^2}}{(r^2 + a^2)^2 \,} \,dr^2-\frac{r^2+\frac{a^4}{a_0^2}}{\sqrt{Q_1Q_5}\,} \,dt^2 + \frac{r^2}{\sqrt{Q_1Q_5}\,}\, dy^2 \ ,
\end{align}
where
\begin{equation}\label{eq:Sigmadef}
    \Sigma\equiv r^2 + a^2 \cos^2\theta\,,\quad  \mathcal{P} \equiv \frac{Q_1Q_5}{\Sigma^2}\bigg[ 1- \frac{a^2b^2}{2\,a_0^2}\, \frac{\sin^2\theta}{r^2+a^2}\bigg]\,.
\end{equation}

The regime in which the CFT state is described by a classical geometry is the one for which both $N$ and $N_b$ are very large numbers. We do, however, have the freedom to choose the ratio $N_b/N$. In the simplest limit, this ratio is small and hence the 3D geometry \eqref{eq.3d100Metric} is a small deformation of global AdS${}_3$ (this can be seen from \eqref{eq.3d100Metric}: when $N_b/N$ and thus $b$ vanish, $a_0=a$ and $ds_3^{\,2}$ becomes AdS${}_3$). To take advantage of this simplification, we can use the small expansion parameter $\mu$ defined by \eqref{eq.alpha} with $n=0$ and \eqref{eq:bN100}:
\begin{align} \label{eq.muba0}
    \sqrt{1-\mu\,} = 1-\frac{N}{N_b} =1-\frac{b^2}{2a_0^2}=\frac{a^2}{a_0^2} \ ,
\end{align}
and perform a perturbative expansion in $\mu$ at fixed $Q_1$, $Q_5$, $R_y\,$, and hence fixed $a_0$. Keeping only the corrections of order $\mu$, the 3D Einstein metric becomes
\begin{align} \label{eq.microExp}
    ds^{\,2}_3 \approx  \frac{\sqrt{Q_1Q_5}\,}{r^2 + a_0^2(1-\mu)}\bigg[1 - \frac{a_0^2}{a_0^2+r^2}\,\mu\bigg]\, dr^2- \frac{r^2 +a_0^2(1 - \mu)\,}{\sqrt{Q_1Q_5}}\, dt^2 + \frac{r^2}{\sqrt{Q_1Q_5}\,}\, dy^2  \ .
\end{align}
The $g_{tt}$ and $g_{yy}$ components of this metric match exactly those of the conical defect metric \eqref{eq.DefectGeoAdS3} with $\tfrac{a}{k}$ replaced with $a_0(1-\mu)^{1/2}$, whereas $g_{rr}$ receives corrections in $\mu$ already at first order. To make certain that this difference is not simply a coordinate artefact, one can compute the Ricci and Kretschmann scalars for the metric \eqref{eq.3d100Metric} to first order in $\mu$
\begin{align}
    \sqrt{Q_1Q_5}\:\mathcal{R} \approx -6 - \frac{2a_0^2(2a_0^2+r^2)}{(a_0^2+r^2)^2}\mu \ \ ,\quad Q_1Q_5\,\mathcal{K} &\approx 12 + \frac{8a_0^2(2a_0^2+r^2)}{(a_0^2+r^2)^2}\mu \ ,
\end{align}
and note that they differ by order $\mu$ terms from the (normalised) conical defect values $\mathcal{R}=-6$ and $\mathcal{K}=12$. Therefore, the conical defect geometry \eqref{eq.DefectGeoAdS3} and the microstate geometry \eqref{eq.3d100Metric} are physically distinct already at first order in $\mu$ and only the latter is dual to a state of the CFT for generic values of $\mu$.

Exploiting the separability of the $(1,0,0)$ family of microstates, one can compute the bulk phase shift in the reduced 3D metric \eqref{eq.3d100Metric} by applying the general formula \eqref{eq.BulkPS3}. This yields
\begin{align}
    \delta_{b} &= 2a_0R_y\,|p_t|\!\int_{r_0}^{\infty}\!dr\,\big(r^2+a^2\big)^{-1}\sqrt{1-\frac{\beta^2}{r^2}\bigg(r^2+\frac{a^4}{a_0^2}\bigg)} \nonumber\\
    &= \pi R_y\,|p_t|\, \abs{\beta}\, \bigg(\!-1+\sqrt{1+\frac{a_0^2}{a^2}\big(\beta^{-2}-1\big)\,}\,\bigg) \ ,
\end{align}
where the radial turning point, obtained by setting to zero~\eqref{eq.NullEOM}, is
\begin{align}\label{eq:turning100}
    r_0 = \frac{a^2}{a_0} \big(\beta^{-2}-1\big)^{-\frac{1}{2}} \ .
\end{align}
Subtracting the phase shift for pure AdS (corresponding to $b=0$) gives
\begin{align} \label{eq.100BulkPS}
    \delta = \delta_{b} - \left.\delta\,\right|_{b=0}= \pi R_y\,|p_t|\,\bigg(\!-1+ \sqrt{\frac{2a_0^2-b^2\beta^2}{2a_0^2-b^2}\,}\,\bigg) \ ,
\end{align}
where we used~\eqref{eq:a0def} to express the result in terms of $a_0$ and $b$. Though the phase shift in \eqref{eq.100BulkPS} is exact in $b$, we will only attempt a CFT interpretation perturbatively in the small $b$ (small $\mu$) limit, describing small deviations from the AdS${}_3$ vacuum. The first two terms in the perturbative expansion of the phase shift for small $\mu$ are
\begin{align} \label{eq.100BulkPSmu}
    \delta \approx \pi R_y\,|p_t|\, \bigg[\frac{\mu}{4}(1-\beta^2) + \frac{\mu^2}{32}(1-\beta^2)(5+\beta^2) + \cdots \bigg] \ .
\end{align}
It is noted that the above expansion is in small $\mu$ but fixed impact parameter $\beta$ and hence it also applies to the regime of $\beta$ small, in which the geodesic explores the region deep inside the bulk. In the next section we will give a CFT derivation of the order $\mu$ term in \eqref{eq.100BulkPSmu}. We conclude the bulk analysis with a comment: the phase shift is expected to be dominated by the graviton exchange which, in the limit of large $s$ and $L$~\eqref{eq.ptpyRelations}, implies a behaviour of the form $\delta \sim s\,e^{-L}$ for a 3D bulk (see for example \cite{Kulaxizi:2017ixa}). Taking the large $L$ (or equivalently the $\beta\to 1$) expansion of the phase shift \eqref{eq.100BulkPS} gives
\begin{align}
    \delta \approx \pi R_y \,s\, e^{-L} \frac{b^2}{2 a_0^2}\bigg(1 -\frac{b^2}{2 a_0^2}\bigg)^{\!-1} \ ,
\end{align}
consistent with the expected generic behaviour mentioned above. This regime describes geodesics with large impact parameter, probing only a shallow region inside the bulk, though we will later check explicitly that the full phase shift is determined by the graviton exchange.

\subsection{The CFT description}
\label{ssec:CFT}

With the aim of reproducing the phase shift \eqref{eq.100BulkPSmu} from a purely CFT computation, we consider the four-point correlation function $C=\langle\mathcal{O}_H\mathcal{O}_L\bar{\mathcal{O}}_L\bar{\mathcal{O}}_H\rangle$ in the supergravity regime. Again both $\mathcal{O}_L=\mathcal{O}^{\mathrm{bos}}$ and $\mathcal{O}^{\mathrm{fer}}$ are considered for the light operator, while the heavy operator is $\mathcal{O}_H=(\mathcal{O}^{\mathrm{fer}})^{N_b}$, dual to the $(1,0,0)$ geometry with reduced metric \eqref{eq.3d100Metric}. In the case that the CPO's $\mathcal{O}^{\mathrm{fer}}$ appearing in the light and the heavy operators belong to different 6D multiplets\footnote{When all operators in the correlator descend from the same 6D multiplet, the HHLL correlator contains extra contributions that were not computed in \cite{Galliani:2017jlg,Bombini:2017sge}. The LLLL version of this correlator was derived in \cite{Giusto:2018ovt} and it will be analysed in Section~\ref{ssec:light100}.}, the correlator $C^{\mathrm{fer}}$ -- containing the light operator $\mathcal{O}^{\mathrm{fer}}$ -- was computed in the supergravity limit at first order in $\tfrac{b^2}{a_0^2}$ in~\cite{Galliani:2017jlg} and its completion to all orders in $\tfrac{b^2}{a_0^2}$ was found in the form of a double sum in \cite{Bombini:2017sge}. Here we need only the 
$O(\tfrac{b^2}{a_0^2})$ result, which in the NSNS sector reads
\begin{align} \label{eq.Cfer2}
    C^{\text{fer}} \approx \frac{1}{\abs{1-z}^2} + \frac{b^2}{2a_0^2} \bigg[\frac{N}{2} - \frac{1}{\abs{1-z}^2}+\frac{2}{\pi} |z|^2 \hat{D}_{1122} \bigg]\ ,
\end{align}
where\footnote{Here we follow the conventions of~\cite{Galliani:2017jlg,Bombini:2019vnc}.}
\begin{equation} \label{eq:d1122}
    \frac{2}{\pi} |z|^2 \hat{D}_{1122} = - \frac{4i\,\abs{z}^2}{(z-\zb)^2} \bigg(\frac{z+\zb}{z-\zb}\, D_2(z,\zb) + \frac{\log\abs{1-z}^2}{2i}  + \frac{z+\zb-2\abs{z}^2}{4i\,\abs{1-z}^2}\log\abs{z}^2 \bigg)
\end{equation}
with $D_2$ being the Bloch-Wigner function given by
\begin{align} \label{eq.BlochWigner}
    D_2(z,\zb)
    &= \frac{1}{2i}\bigg[ \Li_2(z) - \Li_2(\zb) + \log\!\abs{z}\log\Big(\frac{1-z}{1-\zb}\Big) \bigg] \ .
\end{align}
Due to the same supersymmetric Ward identity~\eqref{eq.DefectCorrbos} used in the previous section, one can easily obtain the correlator $C^{\text{bos}}$ involving the bosonic light operator from $C^{\text{bos}} = \pd\pdb\, \big[C^{\text{fer}}\big]$. Performing the analytic continuation~\eqref{eq.ReggeLimit} and extracting the imaginary part of the correlator $C^{\mathrm{bos}}$ at first order in $\frac{b^2}{a_0^2} \approx \mu$ we obtain
\begin{align}\label{eq.DirectChannelPole100}
    \left. \text{Im}\,C^{\text{bos}}_{\lcirclearrowright}\right|_{\mu^1} &\approx \frac{2 \pi}{\sigma^4 \eta^2}\left( \frac{1-8\eta + 8\eta^3 - \eta^4 - 12\,\eta^2\log\eta\,}{\sigma(1-\eta)^5} + O(\sigma^{0})\right) \ ,
\end{align}
where the parametrisation \eqref{eq.zzbParam} is used to go to the Regge limit. It is noted that the power of $\sigma$ in \eqref{eq.DirectChannelPole100} again contains the contribution of the $\abs{1-z}^{-4h_2}$ prefactor in \eqref{eq.CorrExp} as well as of the leading Regge term of the exchanged operator. Further taking the limit $\eta\to0$ of \eqref{eq.DirectChannelPole100} selects the exchanged operator of minimal $\hb$, i.e. the stress tensor: its contribution is captured by the global block with $h=2$, $\hb=0$ and is given by $\text{Im}\,C^{\text{bos}}_{\lcirclearrowright} \approx \frac{2\pi}{\eta^2\sigma^5}\,\mu\,$.

As was done for the case of the conical defect, one can try to match the higher order terms in the $\eta$ expansion of~\eqref{eq.DirectChannelPole100} with the spin-$2$ operator blocks corresponding to the exchange of spin-$2$ double-trace operators $\mathcal{O}_{22}$. A new feature of~\eqref{eq.DirectChannelPole100} is the appearance of a term proportional to $\log\eta$ related to the anomalous dimensions\footnote{In this discussion we use the $\bar\delta$ quantities introduced just below~\eqref{eq:VdV} rather than the more common $\gamma$'s. The $\bar\delta$'s include the couplings $c^{\,2}_{\!(0)}$ which bring a dependence on $N_b$ -- see the comments before~\eqref{eq:finv}.} of the non-BPS double-trace operators $\mathcal{O}_{22}=\mathcal{O}_{LL}\equiv\bar{\mathcal{O}}_{\!L} \pd^m \pdb^{\mb}\mathcal{O}_{\!L}\,$. This can also be seen from the direct channel Euclidean decomposition where terms containing $\log|1-z|^2$ appear, from which we can extract the CFT data of~\eqref{eq:finv}:
\begin{equation} \label{eq.Edelb0}
\begin{aligned}
    & \bar\delta({0,0})=\frac{1}{30}\;,\quad \bar\delta({1,1})=\frac{1}{42}\;,\quad\bar\delta(2,2) = \frac{6930}{1102500}\;,\\
    & \bar\delta({2,0})=-\frac{3}{350}\;,\quad \bar\delta({3,1})=-\frac{2}{735} \;,\quad \bar\delta({4,2})=-\frac{462000}{896464800}\;\ldots\;\;,
\end{aligned}
\end{equation}
while all contributions from operators with odd spin and operators with spin higher than two vanish. These Euclidean results can again be checked by comparing with the expansion of the $\log\eta$ term in~\eqref{eq.DirectChannelPole100}
\begin{equation} \label{eq:lnetaexp1}
    \frac{-12\eta^2}{(1-\eta)^5} \approx -12 \eta^2 - 60 \eta^3 - 180 \eta^4\;,
\end{equation}
which agrees with the spin-2 contributions in~\eqref{eq.Edelb0} after multiplication by the factor present in~\eqref{eq.GlobalRegge}. As an example, for $m=2,3,4\ldots$ we have
\begin{equation} \label{eq:421}
  \frac{\Gamma(2m+4)\,\Gamma(2m+3)}{\Gamma^4(m+2)}\, \frac{\mu}{2}  \,\bar\delta({m,m-2}) \to \mu\,(-12,-60,-180,\ldots)\;.
\end{equation}
A similar check can also be performed for the terms in~\eqref{eq.DirectChannelPole100} that are not proportional to $\log\eta\,$: as for the conical defect case in~\eqref{eq.EopeCD}, these contributions should be compared with couplings $c_{\!(1)}^{\,2}(m,\bar{m})$ in~\eqref{eq:finv}. In Section~\ref{ssec:light100} we will discuss in more detail a similar comparison for the LLLL correlator with all operators in the same 6D multiplet -- the interest in this case is due to its Euclidean decomposition involving also operators of spin larger than two.

We now consider the order $\mu$ Regge crossing equations \eqref{eq.mu1CrossingH}. On the gravity side, we can read off the anomalous dimensions $\Gamma^{(1)}_{m,\mb}$ from the leading eikonal~\eqref{eq.100BulkPSmu} by using \eqref{eq.AnomDimPSmu} and the identifications \eqref{eq.sLRelations}
\begin{align} \label{eq.BulkPSmu}
    \delta^{(1)}_{\mathrm{bulk}} = \frac{\pi}{4}R_y\,|p_t|\,(1-\beta^2) = \pi\frac{m\,\mb}{m+\mb} \ ,
\end{align}
obtaining
\begin{equation} \label{eq.AnomDim100}
    \Gamma^{(1)}_{m,\mb} \approx -\frac{m\,\mb}{(m+\mb)} \ .
\end{equation}
As discussed in~\cite{Li:2017lmh,Karlsson:2019qfi}, it is also possible to use the leading small $\eta$ behaviour of \eqref{eq.DirectChannelPole100} along with the OPE coefficients \eqref{eq.OPEcoeffHR} (with $h_2=1$) to fix the anomalous dimensions $\Gamma^{(1)}_{m,\mb}$. As an example of how this approach works, we start from an ansatz for $\Gamma^{(1)}_{m,\mb}$ in the limit of large $m$ and $\mb$ (that is inspired by, but more general than, the one in \eqref{eq.AnomDim100})
\begin{align} \label{eq.100gammaAnsatz}
    \Gamma^{(1)}_{m,\mb} \approx \frac{A\, m^a\mb^a}{(m+\mb)^c} \ ,
\end{align}
and show that the bootstrap constraints require $a=c=-A=1$, as predicted by the gravity computation. As a first step we approximate the sums in \eqref{eq.mu1CrossingH} by integrals
\begin{align} \label{eq.100CrossIntegral}
    \left. \text{Im}\,C^{\mathrm{bos}}_{\!\lcirclearrowright}\right|_{\mu^1} &\approx -\pi A\! \int_0^{\infty}\!\!\!\!\int_0^{\infty}\!d\mb\,dm\, \frac{\,m^{a+1}\mb^{a+1}}{(m+\mb)^c}\, z^m\zb^{\mb} \equiv -\pi A\, I_{a,c}(z,\zb) \ .
\end{align}
This integral is discussed in Appendix~\ref{App1}: by using~\eqref{eq:IHabres} and then focusing on the leading contribution for small $\sigma$ we obtain
\begin{align} \label{eq.ImCrossing100}
  \left. \text{Im}\;C^{\mathrm{bos}}_{\!\lcirclearrowright}\right|_{\mu^1} &\approx -\pi A\, \frac{\Gamma^2(a+2)\,\Gamma(2a+4-c)}{\Gamma(2a+4)} \,\eta^{c-a-2}\sigma^{c-2a-4}\, {}_2F_1(a+2,c;2a+4;1-\eta)\;.
\end{align}
Demanding that the leading small $\eta$ contribution reproduces that of~\eqref{eq.DirectChannelPole100} fixes the ansatz parameters to $a=c=-A=1$. Substituting these values back into the full Regge result for the cross channel~\eqref{eq.ImCrossing100} reproduces exactly the direct channel expression \eqref{eq.DirectChannelPole100} for any $\eta$. This implies that the anomalous dimensions of the $\mathcal{O}_{\!H\!L}$ operators in the Regge limit are given by the expression in~\eqref{eq.AnomDim100}. In the lightcone OPE limit $m\gg \mb\gg 1$ these anomalous dimensions reduce to $\Gamma^{(1)}_{m,\mb} \approx - \mb$, the result obtained from the conical defect geometry of section \ref{sec:Defect} and in \cite{Kulaxizi:2018dxo} from considering the above CFT analysis for the Virasoro vacuum block. This match is unsurprising since it was shown in \cite{Bombini:2017sge} that the correlator \eqref{eq.Cfer2} at order $b^2$ reduces to the Virasoro block of the identity in the lightcone OPE limit. Finally, the anomalous dimensions~\eqref{eq.AnomDim100} can be confirmed by a Euclidean block decomposition of the correlator $C^{\mathrm{bos}}$ in the cross channel, from which one can extract the anomalous dimensions at first order in $\mu$ but for finite values of $m$ and $\mb$. With the approximation \eqref{eq.HypergeoApprox} for the blocks, the anomalous dimensions $\Gamma^{(1)}_{m,\mb}$ are the coefficients of the $z^m {\bar z}^{\mb} \log|z|^2$ terms in the $z,{\bar z} \to 0$ expansion of the correlator $C^{\text{bos}}|_{\mu^1}$ divided by $C^{\,2}_{\!(0)}(m,\mb)$. By looking at the first few terms, it is simple to infer that
\begin{equation} \label{eq.AnomDim100exact}
    \Gamma^{(1)}_{m,\mb} = - \frac{(m+1)\,(\mb+1)}{(m+\mb+2)} \ , 
\end{equation}
which agrees with \eqref{eq.AnomDim100} in the large $(m,\mb)$ limit. We have checked that \eqref{eq.AnomDim100exact} correctly reproduces the anomalous dimensions up to order 10 in the Euclidean expansion.

Of course, a similar analysis can also be carried out in much the same fashion for the four-point function with light operator $\mathcal{O}_L=\mathcal{O}^{\mathrm{fer}}$, given in \eqref{eq.Cfer2}. Performing the analytic continuation to the Regge limit gives the leading term in small $\sigma$ as
\begin{align} \label{eq.CferReggemu}
    \left.\text{Im}\,C^{\text{fer}}_{\lcirclearrowright}\right|_{\mu} \approx \frac{\pi}{\eta\sigma^2} \left(\frac{1-\eta^2 + 2\eta\log{\eta}}{(1-\eta)^3\,\sigma} + O(\sigma^{0})\right) \ ,
\end{align}
where the factor of $\eta^{-1}\sigma^{-2}$ comes from the usual prefactor $(1-z)^{-2h_2}(1-\bar{z})^{-2\hb_2}$ in the correlator. We note that, as was the case for the conical defect correlators, the Regge limit results in~\eqref{eq.CferReggemu} and~\eqref{eq.DirectChannelPole100} are directly related by~\eqref{eq:zbse}. Another explicit check we can perform in this case is that the Regge limit is dominated by the highest spin field exchanged between the light and heavy operators. In the supergravity approximation being used, this is just the graviton. For the case of~\eqref{eq.CferReggemu}, we can use the results of~\cite{Rastelli:2019gtj} where the contribution of the Witten diagram describing graviton exchange was calculated for the correlator involving four light operators of dimension $(h_L,\hb_L)=(1/2,1/2)$. Since the small $\tfrac{b^2}{2a_0^2}$ limit of the HHLL correlator smoothly reproduces the light one~\cite{Giusto:2019pxc}, we can obtain the first order contribution from the graviton exchange simply by multiplying the result of~\cite{Rastelli:2019gtj} by $\tfrac{b^2}{2a_0^2}$ to get
\begin{equation} \label{eq.gexcOf}
    C^{\text{fer}}_{\text{grav}} = \frac{b^2}{2 a_0^2} \left[\frac{2}{\pi} (z+\bar{z}) \hat{D}_{1122}-\frac{1}{|1-z|^2}\right] \ ,
\end{equation} 
where $\hat{D}_{1122}$ was defined in \eqref{eq:d1122}. By performing the usual analytic continuation relevant for the Regge limit on~\eqref{eq.gexcOf} one obtains, as expected, the result~\eqref{eq.CferReggemu} derived from the full amplitude.

The cross channel calculation follows that of the bosonic case closely: using the same Regge limit ansatz \eqref{eq.100gammaAnsatz} and the order $\mu^0$ OPE coefficients \eqref{eq.OPEcoeffHR}, now with $h_2=\bar{h}_2=1/2$, \eqref{eq.mu1CrossingH} gives
\begin{equation}
    \left.\text{Im}\;C^{\text{fer}}_{\lcirclearrowright}\right|_{\mu} 
    \approx -\pi A \int_0^{\infty}\!\!\!\!\int_0^{\infty} dmd\mb\, \frac{m^a\mb^a}{(m+\mb)^c}\, z^{m}\zb^{\mb} = -\pi I_{a-1,c}(z,\bar{z})\;.
\end{equation}
Again by using \eqref{eq:IHabres} in the leading small $\sigma$ approximation, the choice $a=c=-A=1$ is necessary to reproduce \eqref{eq.CferReggemu} exactly. Therefore, the anomalous dimensions at order $\mu$ from the fermionic correlator appear to be the same as from the bosonic one -- thus from \eqref{eq.AnomDimPSmu}, the first order bulk phase shifts will also match. This is an explicit check of the universality of the Regge limit since the bulk analysis is independent of the nature of the probe used. 

We conclude this analysis by rederiving the anomalous dimensions~\eqref{eq.100gammaAnsatz} in yet one further way. As mentioned after~\eqref{eq.DTDims}, these anomalous dimensions describe the binding energy of a non-BPS bound state between the original heavy operator and the probe. From the bulk point of view, these binding energies can be derived by studying the equation of motion of the supergravity state dual to the light probe when propagating in the background dual to the heavy operator. In~\cite{Kulaxizi:2018dxo}, the case of a bulk scalar propagating in the asymptotically AdS$_{d+1}$ Schwarzschild geometry was studied up to second order. In the case discussed here, we can still focus on a minimally coupled scalar -- dual to the operator $\mathcal{O}^{\mathrm{bos}}$ -- but in the geometry relevant for the heavy state discussed at the beginning of this section. The energies of the bound states in this geometry were derived exactly in $b^2$ in~\cite{Bombini:2017sge}; see\footnote{The parameters $l$ and $n$ appearing in that equation are the spin $l=m-\mb$ and twist $n=\min(m,\mb)+1$.} Eq.~(3.43) of that reference, which in our notation reads
\begin{align} \label{eq.BindingEnergy}
    \omega_n &= \frac{a}{a_0}\sqrt{ (m+\mb+2)^2 + (m-\mb)^2\frac{b^2}{2a^2}} \nonumber\\
    &\approx (m+\mb) - \frac{m\mb}{m+\mb}\mu - \frac{m\mb(m^2 + 4m\mb+\mb^2)}{4(m+\mb)^3}\mu^2 + O(\mu^3) \ ,
\end{align}
where in the second line we performed both the small $\mu$ and the large $m,\mb$ expansions. At first order in $\mu$ this matches precisely~\eqref{eq.AnomDim100}. It is also noted that, by keeping $m$ and $\mb$ exact while expanding the first line of \eqref{eq.BindingEnergy} in $\mu$, the finite shifts of~\eqref{eq.AnomDim100exact} are reproduced. By using the result above, it is straightforward also to check the relation between anomalous dimensions and the phase shift at second order from~\cite{Karlsson:2019qfi}. The second-order version of~\eqref{eq.AnomDimPSmu} reads
\begin{equation} \label{eq.AnomDimPSmu2o}
 \Gamma_{m,\mb}^{(2)} \approx -\frac{\delta^{(2)}}{\pi} + \frac{1}{2} \frac{\delta^{(1)}}{\pi}(\partial_m +\partial_{\bar{m}})  \frac{\delta^{(1)}}{\pi}  \quad \mathrm{for}\quad m,\mb\gg 1\ .
\end{equation}
It is straightforward to check that this identity is satisfied if the ${\cal O}(\mu^2)$ term of~\eqref{eq.BindingEnergy} is used for the left hand side, while the right hand side is calculated using~\eqref{eq.100BulkPSmu} and the identifications~\eqref{eq.sLRelations}.

\subsection{Light case}
\label{ssec:light100}

In the preceding section, correlators involving the heavy operator $\mathcal{O}_H=(\mathcal{O}^{\mathrm{fer}})^{N_b}$ were considered in the scaling limit $N_b\sim N\to\infty$. This amounts to taking the number of non-trivial single-particle constituents in the heavy state to be of order $N$ (to have a backreaction on the dual geometry) but small enough for $\tfrac{N_b}{N}=\tfrac{b^2}{2a_0^2}$ to be a meaningful expansion parameter. Alternatively, it is possible to consider these correlators in the scaling limit $N\to\infty$ with $N_b$ fixed. This implies that the dimension of the `heavy' operator, which scales as $h^{[0]}_H\sim N_b\sim N\tfrac{b^2}{a_0^2}\,$, is no longer of order $N$. In the bulk it is therefore no longer dual to a semi-classical geometry that differs from pure AdS${}_3\times\,$S${}^3$ and in the CFT analysis the approximation \eqref{eq.HypergeoApprox} is no longer valid. However, the $N_b\to 1$ limit of the HHLL correlator reproduces the LLLL correlator~\cite{Giusto:2019pxc}. Then for instance, $C^{\mathrm{bos}}$ at order $b^2$ in the light scaling limit is equal to the following LLLL four-point function
\begin{align}
    \left. C^{\mathrm{bos}}_L\right|_{b^2} = \langle \mathcal{O}^{\mathrm{fer}}(\infty) \mathcal{O}^{\mathrm{bos}}(1) \bar{\mathcal{O}}^{\mathrm{bos}}(z,\zb) \bar{\mathcal{O}}^{\mathrm{fer}}(0) \rangle \ .
\end{align}
However, even if the analytic form of the LLLL correlator is identical to that of the HHLL correlator at order $\mu$, the CFT data obtained in the Regge limit are different. Here we briefly discuss the LLLL analysis following \cite{Li:2017lmh}: the key difference with the HHLL case is that we now need to use the approximation for the conformal blocks in terms of Bessel functions~\eqref{eq.BlockApproxL}. As before, the Regge limit crossing equations \eqref{eq.mu1Crossing2} can be used to solve for the anomalous dimensions of the double-trace operators
\begin{align}\label{eq:OLL'}
\mathcal{O}_{LL'} \equiv\ :\!\mathcal{O}^{\text{fer}}\pd^m\pdb^{\mb}\mathcal{O}^{\text{bos}}\!: \ ,
\end{align}
exchanged in the cross channel. In the Regge limit, in which operators with large $m,\mb$ dominate, the OPE coefficients \eqref{eq.OPEcoeff} with external operator dimensions $2h_1=h_2=1$ reduce to
\begin{equation} \label{eq.OPECoeffLR}
    C^{\,2}_{\!(0)} = \frac{\Gamma^2(2+m)\,\Gamma^2(2+\mb)}{\Gamma(2+2m)\,\Gamma(2+2\mb)} \approx \frac{\pi}{4}\,2^{-2(m+\mb)}(m\mb)^{\tfrac{3}{2}} \ .
\end{equation}
Using \eqref{eq.BlockApproxL} and \eqref{eq.OPECoeffLR} in the first order Regge crossing equations \eqref{eq.mu1Crossing2} (for the LLLL case, {\it i.e.} with $\gamma^{(1)}$ instead of $\Gamma^{(1)}$) gives
\begin{align} \label{eq.ImReggeCrossingL2}
    \text{Im}\left. C^{\text{bos}}_2\right|_{\lcirclearrowright} 
    &\approx -16\pi\, \abs{1-z}^{-1}\!\! \int_{0}^{\infty}\!\!dm\!\!\int_{0}^{m}\!\!d\mb\, (m\mb)^{2} \gamma^{(1)}_{m,\mb}\bigg[K_1\Big(2m\sqrt{1-z\,}\Big)K_1\Big(2\mb\sqrt{1-\zb\,}\Big) \nonumber\\
    &\qquad\qquad\qquad\qquad\qquad + K_1\Big(2\mb\sqrt{1-z\,}\Big)K_1\Big(2m\sqrt{1-\zb\,}\Big)\bigg] \ ,
\end{align}
where we took the large $m,\mb$ limit so that the sums can be substituted by integrals and the Bessel functions approximated using
\begin{align} \label{eq.BesselKapprox}
    K_1\Big( 2\hat{z} + 3\sqrt{1-z\,}\Big) \approx K_1\big(2\hat{z}\big) + O\big(\sqrt{1-z\,}\,\big) \ ,
\end{align}
where $\hat{z}\approx m\sqrt{1-z\,}$ is kept fixed as $m\to\infty$. For future convenience, we split the integral into two separate regions and exploited the fact that the anomalous dimensions are invariant under the exchange $m\leftrightarrow \mb$, since all external states are left/right symmetric. Using an ansatz for the leading large $m,\mb$ anomalous dimensions of the form
\begin{equation} \label{eq.AnomDimAnL}
    \gamma^{(1)}_{m,\mb} = A\, (\max(m,\mb))^{a_1} (\min(m,\mb))^{a_2} \ ,
\end{equation}
the two types of integrals in \eqref{eq.ImReggeCrossingL2} are
\begin{align}
    I_1(a_1,a_2,b) &\equiv \int_{0}^{\infty}\!\!dm\!\!\int_{0}^{m}\!\!d\mb\, m^{2+a_1}\mb^{2+a_2}K_b\Big(2\mb\sqrt{1-z\,}\Big)K_b\Big(2m\sqrt{1-\zb\,}\Big)\nonumber\\
    I_2(a_1,a_2,b) &\equiv \int_{0}^{\infty}\!\!dm\!\!\int_{0}^{m}\!\!d\mb\, m^{2+a_1}\mb^{2+a_2}K_b\Big(2m\sqrt{1-z\,}\Big)K_b\Big(2\mb\sqrt{1-\zb\,}\Big) \ .
\end{align}
Solving these integrals as shown in appendix \ref{App2}, the leading order part of \eqref{eq.ImReggeCrossingL2} in small $\sigma$ is given by
\begin{align} \label{eq.ImReggeCrossingL4}
    \left.\text{Im}\,C^{\text{bos}}_L\right|_{\lcirclearrowright} &\approx -A\,\pi\, \sigma^{-4-\frac{1}{2}(a_1+a_2)}\eta^{-1}\bigg[ \!\left.G^{\,2,3}_{3,3}\,\bigg(\eta\;\right|\!\!\!\!\!\!\!
    \begin{array}{cc}
         & -\tfrac{1}{2}(a_1+a_2)-2,-\tfrac{1}{2}(a_1+a_2)-1,-\frac{a_2}{2} \\
         & 1,0,-\frac{a_2}{2}-1
    \end{array}\! \bigg) \nonumber\\
    &\qquad\qquad\ \  + \eta^{-\frac{a_1}{2}-1} \left.G^{\,2,3}_{3,3}\,\bigg(\frac{1}{\eta}\,\right|\!\!\!\!\!\!
    \begin{array}{cc}
         & -\tfrac{1}{2}(a_1+a_2)-2,-\tfrac{1}{2}(a_1+a_2)-1,-\frac{a_2}{2} \\
         & 1,0,-\frac{a_2}{2}-1
    \end{array}\! \bigg)\, \bigg] \ ,
\end{align}
where $G^{\,m,n}_{p,q}$ is the Meijer G-function (defined in Eq.~\eqref{eq.MeijerGdef}). Expanding in small $\eta$ and matching the powers of $\sigma$ and $\eta$ of the leading order term to the contribution of the stress tensor fixes $A=-1$, $a_1=0$ and $a_2=2$. Inserting these values of the ansatz parameters in \eqref{eq.ImReggeCrossingL4} gives precisely \eqref{eq.DirectChannelPole100} and so the anomalous dimensions solving the crossing equations in the Regge limit are
\begin{align} \label{eq.LLLLanomDim}
    \gamma^{(1)}_{m,\mb} \approx -(\min(m,\mb))^2 \ .
\end{align}
Therefore, the anomalous dimensions of the $\mathcal{O}_{LL'}$ operators~\eqref{eq:OLL'} in the Regge limit take a qualitatively different form from their HL counterpart~\eqref{eq.AnomDim100} and agree with the structure expected from the analysis of~\cite{Giusto:2018ovt} (see Eq.~(5.3) of that reference). 

We conclude this section by discussing the Regge limit of another LLLL correlator
\begin{align}\label{eq:G1111ef}
    \langle \mathcal{O}^{\mathrm{fer}}(\infty) \mathcal{O}^{\mathrm{fer}}(1) \bar{\mathcal{O}}^{\mathrm{fer}}(z,\zb) \bar{\mathcal{O}}^{\mathrm{fer}}(0) \rangle \ ,
\end{align}
given in Eq.~(3.10) of~\cite{Giusto:2018ovt}. This example is different from those considered earlier because the CPO's $\mathcal{O}^{\mathrm{fer}}$ in the correlator descend from the same 6D multiplet and, hence, single-trace operators are exchanged also in the cross channel. This implies that in the direct channel, double-trace operators of arbitrarily high spin are exchanged, as can be checked explicitly from the $z,\zb\to1$ Euclidean OPE~\eqref{eq:finv}. For the correlator~\eqref{eq:G1111ef}, the leading direct channel OPE coefficients $c^{\,2}_{\!(0)}$ are
\begin{equation}
c^{\,2}_{\!(0)}(m,\mb)=(-1)^{m+\mb} C^{\,2}_{\!(0)}(m,\mb)\,,
\end{equation}
with $C^{\,2}_{\!(0)}(m,\mb)$ given in \eqref{eq.OPEcoeff}. For the anomalous dimensions of the double-trace operators exchanged in the direct channel one finds
\begin{equation}
  \label{eq:efr0}
  \begin{gathered}
    \bar{\delta}({0,0})= -\frac{5}{6}\;,~~\bar{\delta}({1,0})=-\frac{5}{6}\, c^{\,2}_{\!(0)}(1,0)\;,~~\bar{\delta}({2,0})=-\frac{14}{15}\, c^{\,2}_{\!(0)}({2,0})\;,\\
    \bar{\delta}({m,0})= -c^{\,2}_{\!(0)}(m,0)\quad\ \ \mbox{for }\ m>2\;,
  \end{gathered}
\end{equation}
when focusing on the case $\bar{m}=0$ and
\begin{equation}
  \label{eq:efr1}
  \begin{gathered}
  \bar{\delta}({1,1})=-\frac{61}{30}\, c^{\,2}_{\!(0)}({1,1})\;,~\bar{\delta}({2,1})=-\frac{41}{15}\, c^{\,2}_{\!(0)}({2,1})\;,~\bar{\delta}({3,1})= -\frac{102}{35} c^{\,2}_{\!(0)}({3,1})\;,\\ \bar{\delta}({m,1})=  -3\,c^{\,2}_{\!(0)}({m,1})\quad\ \ \mbox{for }\ m>3\;,
\end{gathered}
\end{equation}
for $\bar{m}=1$. As expected, the data for operators of spin larger than two takes the form
\begin{equation} \label{eq:bargamma}
    \bar{\delta}({m,\bar{m}})=-(n^2+n+1)\, c^{\,2}_{\!(0)}(m,\bar{m})\quad \mathrm{with}\quad n = \mathrm{min}(m,\bar m)~\mbox{ and }~ |m-\mb|>2\;,
\end{equation}
in agreement with~\cite{Giusto:2018ovt}. Clearly in this case one cannot follow the previous approach, of performing the Regge limit on the contribution of each block separately, since this would lead to poles $\sigma^{-a}$ with $a>1$. Such contributions are absent in the Regge limit of the correlator, which is again given by~\eqref{eq.CferReggemu}. We instead first need to resum all contributions with $m>\mb+2$ and then perform the analytic continuation~\eqref{eq.ReggeLimit}. The result of this resummation, made possible by exploiting~\eqref{eq:bargamma}, is an {\em analytic} term around $z=0$ (which does not contribute to the Regge limit) and a contribution equal to the naive extension of~\eqref{eq:bargamma} to $m=\mb+2$. Then as before, the contribution to the $\log\eta$ term of~\eqref{eq.CferReggemu} comes entirely from the operators of spin~$2$ and one can use~\eqref{eq:finv} to relate the Regge limit and anomalous dimensions for these operators, obtaining
\begin{equation} \label{eq:d20checkg2}
    \bar{\delta}(\mb+2,\,\mb) = \frac{(\mb+1) (\mb+2)\,\Gamma^4(\mb+3)}{ \Gamma(2\mb+6)\,\Gamma(2\mb+5)} - (\mb^2+\mb+1) \frac{\Gamma^2(\mb+3)}{\Gamma(2\mb+5)}\frac{\Gamma^2(\mb+1)}{\Gamma(2\mb+1)}\;.
\end{equation}
The first term in this expression encodes the input from the Regge limit and comes from the small $\eta$ expansion of the $\log\eta$ term in~\eqref{eq.CferReggemu}. For example, the results of~\eqref{eq:efr0} and~\eqref{eq:efr1} are reproduced for $\mb=0,1$.

A similar argument can be used to also derive the couplings $c^{\,2}_{\!(1)}(\mb+2,\mb)$. Again one can use the asymptotic result~\cite{Heemskerk:2009pn,Alday:2017gde,Alday:2017vkk} $c^{\,2}_{\!(1)}({m,\bar{m}}) = (\partial_m+\partial_{\bar{m}})\bar{\delta}({m,\bar{m}})$, valid for $|m-\mb|>2$, and the terms without $\log\eta$ in~\eqref{eq.CferReggemu} to find the following explicit expression for the couplings
\begin{equation} \label{eq:C1}
    c^{\,2}_{\!(1)}(\bar{m}+2,\bar{m}) = \frac{(2\bar{m}+3)}{\Gamma} - \,\partial_{\bar{m}}\Big((\bar{m}^2+\bar{m}+1)\,C^{\,2}_{\!(0)}(\bar{m}+2,\bar{m})\Big) - \frac{(\bar{m}+1)(\bar{m}+2)}{\Gamma^{2}}\,\partial_{\bar{m}}\Gamma \ ,
  \end{equation}
  where we defined
\begin{equation}
    \Gamma \equiv \frac{\Gamma(2\bar{m}+6)\,\Gamma(2\bar{m}+5)}{\Gamma^4(\bar{m}+3)} \ .
\end{equation}
The first term in~\eqref{eq:C1} comes from the expansion of the explicit correlator in the Regge limit and the other terms are obtained by rearranging the right-hand side of~\eqref{eq:finv}, thanks to~\eqref{eq:d20checkg2} and the relation for $C^{\,2}_{\!(1)}$ for operators of spin $|m-\mb|>2$. We note that~\eqref{eq:C1} agrees with the couplings obtained for the first few values of $\bar{m}$ from the direct channel Euclidean OPE
\begin{align}
    c^{\,2}_{\!(1)}({2,0}) = \frac{19}{1350}\ \ ,\quad c^{\,2}_{\!(1)}({3,1}) = \frac{4331}{49000}\ \ ,\quad c^{\,2}_{\!(1)}({4,2}) = \frac{520433}{18522000} \ .
\end{align}

\section{A class of three-charge microstate geometries} \label{sec:101}

As an extension to the case of the $(1,0,0)$ geometry considered in section \ref{sec:100Geometry}, it is possible to add momentum charge yielding a class of 3-charge microstate geometries. This can be done so as to preserve the separability of the 6D spacetimes into asymptotically $S^3$ and AdS${}_3$ 3-manifolds -- where the Einstein metric of the latter part is independent of the $S^3$ coordinates. From the CFT perspective, these are 1/4-BPS states obtained by acting $n$ times on the single-particle constituents of the $(1,0,0)$ microstates with the Virasoro generator $L_{-1}$. Each of the new $N_b$ single-particle constituents carries $n$ units of momentum along the S${}^1$ of the CFT and the quantised momentum charge of the full microstate is
\begin{equation}
    n_\mathrm{P} = n\, N_b \ .
\end{equation}
On the gravity side, the number of momentum-carrying strands $N_b$ is controlled by the parameter $b$ according to the same relation \eqref{eq:bN100}, though now with a more general dependence on the expansion parameter $\mu$ (given implicitly in \eqref{eq.alpha})
\begin{align} \label{eq.10nMubRel}
    \frac{N_b}{N}= \frac{b^2}{2 a_0^2} = 2n+1-\sqrt{(2n+1)^2-\mu\,} \ ,
\end{align}
where $a_0$ is defined in Eq.~\eqref{eq:a0def}.

The full 10D geometry describing this $(1,0,n)$ family of microstates can be found for example in \cite{Bena:2016ypk,Bena:2017xbt}. For the purposes of calculating the eikonal, it is again useful to write the 6D part of this solution in the dimensionally-reduced form \eqref{eq.kkansatz}, with $S^3$ metric $G_{\alpha\beta}$, gauge fields $A^\alpha$ and 3D Einstein metric $ds^{\,2}_3$ here given by
\begin{equation}
G_{\theta\theta}=\sqrt{\mathcal{P}\,} \,\Sigma\,,\, G_{\phi\phi}=\frac{Q_1 Q_5}{\sqrt{\mathcal{P}\,} \,\Sigma}\,\sin^2\theta\,,\,\,G_{\psi\psi}=\frac{Q_1 Q_5}{\sqrt{\mathcal{P}\,}\, \Sigma}\,\left[1-\frac{a^2 b^2}{2 a_0^2 (r^2+a^2)}\left(\frac{r^2}{r^2+a^2} \right)^n\right]\,\cos^2\theta\,,
\end{equation}
\begin{equation} 
A^\theta=0\,,\,\, A^\phi = -\frac{a^2}{a_0^2} \frac{dt}{R_y}\,,\,\, A^\psi = -\frac{\frac{a^2}{a_0^2} F_n \frac{dt}{R_y} +\left(1-\frac{b^2}{2a_0^2} \left(\frac{r^2}{r^2+a^2}\right)^n \right) \frac{dy}{R_y}}{1-\frac{a^2 b^2}{2 a_0^2 (r^2+a^2)} \left(\frac{r^2}{r^2+a^2}\right)^n}\,,
\end{equation}
\begin{align} \label{eq.3dMetric101}
    ds_3^{\,2} = \sqrt{Q_1Q_5}\,\frac{r^2+\frac{a^4}{a_0^2}\,(1+F_n)}{(r^2 + a^2)^2 \,} \,dr^2-\frac{r^2+\frac{a^4}{a_0^2}}{\sqrt{Q_1Q_5}\,} \,dt^2 + \frac{r^2}{\sqrt{Q_1Q_5}\,}\, dy^2+\frac{r^2 F_n}{\sqrt{Q_1Q_5}\,}\, (dt+ dy)^2 \ ,
\end{align}
where $\Sigma$ is as defined in \eqref{eq:Sigmadef} and
\begin{equation}
\mathcal{P} \equiv \frac{Q_1Q_5}{\Sigma^2}\bigg[ 1- \frac{a^2b^2}{2\,a_0^2}\frac{\sin^2\theta}{r^2+a^2} \left( \frac{r^2}{r^2+a^2}\right)^n\bigg]\ \ ,\quad F_n(r)\equiv \frac{b^2}{2 a^2}\left[1-\left(\frac{r^2}{r^2+a^2}\right)^n\right] \ .
\end{equation}
The geometry \eqref{eq.3dMetric101} is in general difficult to work with and so for simplicity we focus on the particular case of the $(1,0,1)$ 3-charge microstate geometry. The analysis of the bulk eikonal now follows that outlined in section \ref{sec:BulkPS} by considering null geodesics in the 3D geometry \eqref{eq.3dMetric101} (with $n=1$) that begin and end on the boundary. For the purposes of this paper, it is sufficient to evaluate the phase shift integral perturbatively in $\mu$. Starting from \eqref{eq.BulkPS3}, using the change of variables $x=\tfrac{r}{r_0}$ (removing all $b$ dependence from the integral limits) and expanding in $\mu$ using \eqref{eq.10nMubRel} gives
\begin{align}
    \delta &= \int_0^1\!dx\,\delta_x = \sum_{j=0}^{\infty} \int_0^1\!dx\, \delta^{(j)}_x \mu^j \ ,
\end{align}
with the zeroth and first order integrands
\begin{align}
    \delta^{(0)}_x &= 2 |p_t| R_y\frac{\abs{\beta}(1-\beta^2)\sqrt{1-x^2\,}}{x(1-\beta^2)+\beta^2}\label{eq.zerothPS}\\
    \delta^{(1)}_x &= |p_t| R_y\frac{\abs{\beta}^3(1-\beta^2)(3-2\beta+\beta^2)\sqrt{1-x^2\,}}{6\big(x(1-\beta^2)+\beta^2\big)^2} \label{eq.firstPS}\ .
\end{align}
In deriving these integrands the expansion of the turning point in $\mu$ is used
\begin{align}
    r_0 \approx \frac{a_0\abs{\beta}}{\sqrt{1-\beta^2\,}} - \frac{a_0\abs{\beta}(3-2\beta+\beta^2)}{12\sqrt{1-\beta^2\,}}\mu \ .
\end{align}
The zeroth order phase shift obtained from the integral of \eqref{eq.zerothPS} is just that of global AdS${}_3$, $\delta^{(0)}=\pi R_y |p_t|(1-\abs{\beta})$, whereas at first order one gets from \eqref{eq.firstPS}
\begin{align} \label{eq.PS101}
    \delta^{(1)} = \int_0^1\!dx\,\delta^{(1)}_x = \frac{\pi}{24}R_y\,|p_t| (1-\beta^2)(3 - 2\beta + \beta^2) \ .
\end{align}

We would again like to compare this result for the phase shift with information contained in appropriate HHLL 4-point correlators of the dual CFT. In the heavy regime, the 1/4-BPS operators dual to the family of geometries in \eqref{eq.3dMetric101} have reduced dimensions \eqref{eq.HeavyDimn}, which scale with the central charge. These heavy operators will, for generic values of $N_b$ and $n\neq0$, be mixtures of quasi-primary and descendant parts; only in the light limit $N_b\to1$ will they be pure descendants. In the latter case, and for $n=1$, a Ward identity relates the correlator of primary operators with that containing two primaries and two descendants \cite{Bombini:2019vnc}. Exploiting the equivalence of the LLLL and HHLL correlators at order $\mu$, the same Ward identity can be used to derive the $O(\mu)$ correlator in the $(1,0,1)$ heavy state from that in the $(1,0,0)$ heavy state:
\begin{align} \label{eq.101Corr}
    \left.G_{1,0,1}(z,\zb)\right|_{\mu^1} =\Big[(1-z)^2\partial(z\,\partial) +1\Big]\left.G_{1,0,0}(z,\zb)\right|_{\mu^1} \ .
\end{align}
The relation between the respective full correlators can then be obtained from \eqref{eq.CorrExp} by including the appropriate prefactors. In this section we consider only HHLL correlators containing the light operator $\mathcal{O}^{\mathrm{bos}}$ with dimension $(h_2,\hb_2)=(1,1)$.
The order $\mu^0$ piece of this correlator is dependent solely on the dimension of the light operator used and so is equal to the $(1,0,0)$ case and given by $\abs{1-z}^{-4}$. At first order in $\mu$, performing the analytic continuation to the Regge region and extracting the leading imaginary piece gives
\begin{align} \label{eq.DirectChannelPole101}
    \mathrm{Im}\left.C_{1,0,1}^{\mathrm{bos}}\right|_{\lcirclearrowright,\, \mu} \approx 2 \pi \frac{3 - 42 \eta - 199 \eta^2+ 160 \eta^3 + 69 \eta^4 + 10 \eta^5 - \eta^6 - 12 \eta^2 (13 + 14 \eta + 3 \eta^2)\log\eta}{3(1-\eta)^7\eta^2\sigma^5} \ ,
\end{align}
where the relation \eqref{eq.10nMubRel} with $n=1$ and the parametrisation \eqref{eq.zzbParam} have been used. One can also obtain~\eqref{eq.DirectChannelPole101} from~\eqref{eq.DirectChannelPole100} by rewriting the differential operator in~\eqref{eq.101Corr} in terms of $(\sigma,\eta)$ as done in~\eqref{eq:zbse}.

We now move to analysing the cross channel interpretation of \eqref{eq.DirectChannelPole101}. From \eqref{eq.mu1CrossingH} the contributions from double-trace operators of the schematic form $\mathcal{O}_{\!H}\pd^m\pdb^{\mb}\mathcal{O}_L$ can be resummed; again this is dominated by the operators with large $m$ and $\mb$. The anomalous dimensions $\Gamma^{(1)}_{m,\mb}$ in the Regge limit can be extracted from the bulk phase shift \eqref{eq.PS101}, once again using the relations \eqref{eq.AnomDimPSmu} and \eqref{eq.sLRelations}, giving
\begin{align} \label{eq.101AnomDim}
    \Gamma^{(1)}_{m,\mb} \approx -\frac{\delta^{(1)}}{\pi} = -m\mb\frac{m^2 +2m\mb + 3\mb^2}{3(m+\mb)^3} \ .
\end{align}
We note that these anomalous dimensions are not symmetric under the exchange of $m$ and $\mb$, unlike those in the conical defect and $(1,0,0)$ cases. This is to be expected for the 1/4-BPS 3-charge microstates, in this case the $(1,0,n)$ family, as a consequence of having acted with only holomorphic Virasoro modes on the $(1,0,0)$ state. Resumming these double-trace contributions, with the approximation to the OPE coefficients in~\eqref{eq.OPEcoeffHR}, gives
\begin{align} \label{eq.101Cross}
    \left. \mathrm{Im}\,C^{(1,0,1)}_{\lcirclearrowright}\right|_{\mu} &\approx -\pi\!\int_0^{\infty}\!dm\!\int_0^{\infty}\!d\mb\,C_{\!(0)}^{\,2}(m,\mb)\,\Gamma^{(1)}_{m,\mb}\,z^m\zb^{\mb} \nonumber\\
    &= -\pi\!\int_0^{\infty}\!\!dm\!\!\int_0^m\!d\mb\,C_{\!(0)}^{\,2}(m,\mb)\,\big(\Gamma^{(1)}_{m,\mb}z^m\zb^{\mb}+\Gamma^{(1)}_{\mb,m}z^{\mb}\zb^{m}\big) \nonumber\\
    &= \frac{\pi}{3}\Big( I_{2,2,1}(z,\zb) + I_{2,2,1}(\zb,z) +2I_{2,4,3}(z,\zb) + 2I_{4,2,3}(\zb,z)\Big) \ ,
\end{align}
which gives precisely \eqref{eq.DirectChannelPole101} once expanded in $\sigma$. The final line of the above is written in terms of the integral defined in \eqref{eq.Iabc}, whose solution is given in \eqref{eq.GeneralAsymInt}. This matching of \eqref{eq.101Cross} and \eqref{eq.DirectChannelPole101} demonstrates that the anomalous dimensions \eqref{eq.101AnomDim} obtained from gravity are consistent with the crossing relations.

It is, however, curious that such a matching does occur in the 3-charge case using the above method. As mentioned above, in the heavy scaling regime the $(1,0,1)$ operator will not purely be a quasi-primary ($N_b$ is small compared with $N$ but still macroscopic) and so it appears that both the relation \eqref{eq.AnomDimPSmu} and the decomposition of the correlator used in section \ref{sec:CFT} should not hold. Despite this, it seems that in the heavy limit at least, these differences in the key steps of the CFT analysis are subleading in $1/h_H$.

\section{Summary and outlook}
\label{sec:conclusions}

We have studied the Regge limit of four-point AdS$_3$ correlators in the supergravity approximation. For the most part we have concentrated on HHLL correlators in which one pair of operators corresponds to pure 1/2- or 1/4-BPS states with conformal dimensions $\Delta$ of order $N$, in the large $N$ limit. These heavy operators ($\mathcal{O}_H$) are dual to non-trivial asymptotically AdS gravitational backgrounds, while the light operators ($\mathcal{O}_L$) are described in the Regge regime by null geodesics in these geometries -- from this set up, a phase shift can be computed. To relate this bulk phase shift with the CFT data, we have adopted a perturbative approach in $\Delta/N$, limiting ourselves to the first order. In this limit the HHLL correlators we consider take the same functional form as the LLLL correlators where the pair of heavy operators is replaced by two light CPO's. Despite this, different approximations are appropriate in the analyses for the two regimes $\Delta \sim N$ and $\Delta \sim 1$ (see for instance \eqref{eq.HypergeoApprox} and \eqref{eq.BlockApproxL}). This explains why two different sets of CFT data are extracted from the HHLL and the LLLL correlators. We have verified that the relation \eqref{eq.AnomDimPSmu} between the phase shift and the anomalous dimensions of double-trace operators (those exchanged between a heavy and a light operator) is satisfied in all the examples we have analysed. We have also looked at the bootstrap constraints relating the $\mathcal{O}_H\to \mathcal{O}_L$ and the $\mathcal{O}_L\to \mathcal{O}_L$ channels. This latter channel contains a ``universal sector'' that is completely determined by the Virasoro and R-symmetry algebras of the CFT -- and is thus insensitive to the details of the states appearing in the correlator. Truncating a correlator to this universal contribution, as is often done in the literature, amounts to replacing the pure heavy states by a statistical ensemble characterised by $\Delta$. The correlators in pure states, however, also contain a tower of double-trace operators that are needed for consistency with the bootstrap constraints. An intermediate example is represented by the correlator extracted from the conical defect geometry for generic values of the deficit angle: despite this geometry not being dual to a pure state, the correlator satisfies the bootstrap constraint mentioned above. Finally, as a technical by-product, we show that knowledge of the correlator in the Regge limit is enough to fix the anomalous dimensions and three-point couplings of double-trace operators with spin less than or equal to 2 -- these are not captured by the Lorentzian inversion formula \cite{Caron-Huot:2017vep}. We work out explicitly an example with spin-2 operators. 

Our investigation leaves open a number of possible future developments. Firstly, at a more technical level, it would be useful to explain why the relation \eqref{eq.AnomDimPSmu} between phase shift and anomalous dimensions (that is expected to apply only to correlators of quasi-primary operators) also works for the non-primary state considered in Section~\ref{sec:101}. This question becomes particularly relevant because non-primary operators are the microstates of the D1-D5-P black hole. Secondly, it would be interesting to further analyse the conical defect correlator and understand if it can be distinguished from a correlator in an allowed state of the CFT. The most pressing physical question, however, is whether the Regge limit of pure-state correlators can be used as a tool to study the black hole regime of the CFT. Heavy operators are dual to microstates of a black hole with a regular horizon of finite area only if the parameter $\mu$, defined in \eqref{eq.alpha}, is greater than 1 ($\mu>1$) and thus this requirement is incompatible with the approach of this article -- which is based on the small $\mu$ expansion. For the $(1,0,n)$ subfamily of microstates, this translates into the bound
\begin{equation}
    1 \geq \frac{N_b}{N}> 2n+1 -2 \sqrt{n(n+1)}\,,
\end{equation}
which, in particular, requires $n \ge 1$. While computing the full HHLL correlator requires solving the wave equation in the $(1,0,n)$ geometry \eqref{eq.3dMetric101}, and this is difficult to do exactly (see~\cite{Bena:2019azk,Bena:2020yii}), deriving the bulk phase shift is analytically doable, at least for $n=1$. Hence it would be important to understand precisely how the CFT information is encoded in the bulk phase shift for finite values of $\mu$. We believe that this would provide a useful tool to elucidate the gravitational description of black hole microstates and we hope to be able to make progress on this problem in the near future.

\section*{Acknowledgements}

We would like to thank Massimo Bianchi, Jos\'e Francisco Morales, Nejc \v{C}eplak, Manuela Kulaxizi and Andrei Parnachev for discussions, as well as the organisers of the conference ``Black-Hole Microstructure", where some of the results contained in this article were presented. This work was supported in part by the Science and Technology Facilities Council (STFC) Consolidated Grant ST/P000754/1 {\it String theory, gauge theory \& duality} and by the MIUR-PRIN contract 2017CC72MK003.

\appendix

\section{Cross Channel Heavy Integrals} \label{App1}

In the discussion of the HHLL bootstrap constraints of Section~\ref{sec:100Geometry}, we needed to evaluate
\begin{equation} \label{eq:IHab}
    I_{a,c}(z,\zb) \equiv \int_0^{\infty}\!\!\!\!\int_0^{\infty}\!d\mb\,dm\, \frac{\,m^{a+1}\mb^{a+1}}{(m+\mb)^c}\, z^m\zb^{\mb} \;.
\end{equation}
In order to perform this integral, it is first helpful to decouple the two variables by using a Schwinger parameter $t$ to rewrite the denominator of the integrand. This gives the simpler triple integral
\begin{align}
    I_{a,c}(z,\zb) &= \int_0^{\infty}\!dt\, \frac{t^{\,c-1}}{\Gamma(c)}\int_0^{\infty}\!dm\, m^{a+1} e^{-m(t-\log{z})} \int_0^{\infty}\!d\mb\,\mb^{a+1} e^{-\mb(t-\log{\zb})} \nonumber\\
    &= \frac{\Gamma^2(a+2)}{\Gamma(c)}\int_0^{\infty}\!dt\,t^{c-2a-5}\Big(1-\frac{\log{z}}{t}\Big)^{-a-2}\Big(1-\frac{\log{\zb}}{t}\Big)^{-a-2} \ .
\end{align}
Making a change of variables to $x=\tfrac{\tau}{1+\tau}$ where $t=-\tau\log{\zb}$ and using the integral form of the hypergeometric function
\begin{align}
    {}_2F_1(a,b\,;c\,;z) = \frac{\Gamma(c)}{\Gamma(b)\Gamma(c-b)}\int_0^1\!ds\,s^{b-1}(1-s)^{c-b-1}(1-zs)^{-a} \ ,
\end{align}
the final integral can be performed to give
\begin{equation}
  \label{eq:IHabres}
    I_{a,c}(z,\zb) = \frac{\Gamma^2(a+2)\,\Gamma(2a+4-c)}{\Gamma(2a+4) \,\big(\!-\log{\zb}\big)^{2a+4-c}}\,\bigg(\frac{\log{\zb}}{\log{z}}\bigg)^{a+2} {}_2F_1\bigg(a+2,c;2a+4;1-\frac{\log{\zb}}{\log{z}}\bigg) \ .
\end{equation}

In section \ref{sec:101} the resummation of contributions of the double-trace operators $\{\mathcal{O}_{\!H\!L}\}$ requires the computation of integrals of the following, more general, kind
\begin{align} \label{eq.Iabc}
    I_{a,b,c}(z,\zb) \equiv \int_0^{\infty}\!dm\!\int_0^m\!d\mb\, \frac{m^a\mb^b}{(m+\mb)^c}\,z^m \zb^{\mb} \ .
\end{align}
We now derive the result of this integral. Making a change of variables in the $\mb$ integral to $x=\frac{\mb}{m}$ gives
\begin{align}
    I_{a,b,c}(z,\zb) &\equiv \int_0^{\infty}\!dm\, z^m\, m^{a+b-c+1}\!\int_0^1\!dx\, \frac{x^b\, \zb^{\,mx}}{(1+x)^c} \nonumber\\
    &= \int_0^1\!dx\, \frac{x^b}{(1+x)^c} \int_0^{\infty}\!dm\, m^{a+b-c+1} e^{m(x\log\zb + \log{z})} \ .
\end{align}
Using the integral representation of the gamma function 
\begin{align}
    \Gamma(c) = \int_0^{\infty}\!ds\, s^{c-1}e^{-s} \ ,
\end{align}
with $s=-t\,(x\log\zb + \log{z})$ gives
\begin{align} \label{eq.int123}
    I_{a,b,c}(z,\zb) = \frac{\Gamma(a+b+2-c)}{\big(  \!\!-\!\log{z}\big)^{a+b+2-c}}\!\int_0^1\!dx\,x^b\, (1+x)^{-c}\,\bigg( 1+ x\,\frac{\log\zb}{\log{z}} \,\bigg)^{c-a-b-2} \ .
\end{align}
Now with the use of the integral
\begin{align}
    \int_0^1\!dx\,x^{\lambda-1}\,(1-x)^{\mu-1}\,(1-ux)^{-\rho}\,(1-vx)^{-\sigma} = B(\mu,\lambda) \,F_1(\lambda\,;\rho,\sigma\,;\lambda+\mu\,;u,v) \ ,
\end{align}
with $\mu=1$ and $B(\mu,\lambda)$ the Euler beta function, \eqref{eq.int123} becomes
\begin{align} \label{eq.GeneralAsymInt}
    I_{a,b,c}(z,\zb) = \frac{\Gamma(a+b+2-c)}{(1+b)\big(  \!\!-\!\log{z}\big)^{a+b+2-c}}\, F_1\bigg(\!b+1\,;c,a+b+2-c\,;2+b\,;-1,-\frac{\log\zb}{\log{z}}\bigg) \ .
\end{align}
Here $F_1$ is an Appell hypergeometric function with series representation
\begin{align}
    F_1(a;b_1,b_2;c\,;x,y) = \sum_{m,n=0}^{\infty} \frac{\poc{a}{m+n}\poc{b_1}{m}\poc{b_2}{n}}{m!\,n!\,\poc{c}{m+n}}\,x^m y^n \ .
\end{align}
As a check, one can verify that setting $b=a$ in~\eqref{eq.GeneralAsymInt} yields the previous result~\eqref{eq:IHabres}.

\section{Cross Channel Light Integrals} \label{App2}

In solving the Regge limit crossing equations at first order in the $\tfrac{1}{N}$ expansion for the case of all operators being light, the following equation is to be solved
\begin{equation} \label{eq.ImReggeCrossingL3}
    \text{Im}\left. C^{\text{bos}}_2\right|_{\lcirclearrowright} \approx -16\pi A\, \abs{1-z}^{-1} \Big[ I_2(a_1,a_2,1) + I_1(a_1,a_2,1)\Big] \ ,
\end{equation}
where the two integrals required are
\begin{align}
     I_1(a_1,a_2,b) &\equiv \int_{0}^{\infty}\!\!dm\!\!\int_{0}^{m}\!\!d\mb\, m^{2+a_1}\mb^{2+a_2}K_b\Big(2\mb\sqrt{1-z\,}\Big)K_b\Big(2m\sqrt{1-\zb\,}\Big)\nonumber\\
    I_2(a_1,a_2,b) &\equiv \int_{0}^{\infty}\!\!dm\!\!\int_{0}^{m}\!\!d\mb\, m^{2+a_1}\mb^{2+a_2}K_b\Big(2m\sqrt{1-z\,}\Big)K_b\Big(2\mb\sqrt{1-\zb\,}\Big) \ .
\end{align}
Focusing on $I_2$ first, with $b=1$, performing a change of variables in the $\mb$ integral using $\sqrt{x\,}=\tfrac{\mb}{m}$ gives
\begin{align} \label{eq.I2}
    I_2(a_1,a_2,1) &= \frac{1}{2}\int_0^{\infty}\!dm\,m^{5+a_1+a_2}K_1\Big(2m\sqrt{1-z\,}\Big)\int_0^1\!dx\, x^{\frac{1}{2}(1+a_2)} K_1\Big(2m\sqrt{x\,}\,\sqrt{1-\zb\,}\Big) \nonumber\\
    &= \frac{1}{4}(1-\zb)^{-\frac{1}{2}}\!\int_0^{\infty}\!dm\, m^{4+a_1+a_2}K_1\Big(2m\sqrt{1-z\,}\Big) \left.G^{\,2,1}_{1,3}\,\bigg(m^2(1-\zb)\,\right|\!\!\!\!\!\!
    \begin{array}{cc}
        & -\frac{a_2}{2} \\
        & 1,0,-\frac{a_2}{2} -1
    \end{array}\!\bigg) \ ,
\end{align}
where in the second line the following integral from Eq.~(6.592.2) of \cite{Gradshteyn:2007} was used
\begin{align}
    \int_0^1\!dx\,x^{\lambda}(1-x)^{\mu-1} K_{\nu}\big(a\sqrt{x\,}\big) = \frac{2^{\nu-1}}{a^{\nu}}\,\Gamma(\mu)\,\left.G^{\,2,1}_{1,3}\,\bigg(\frac{a^2}{2}\,\right|\!\!\!\!\!\!
    \begin{array}{cc}
        & \frac{\nu}{2}-\lambda \\
        & \nu,0,\frac{\nu}{2}-\lambda-\mu
    \end{array}\!\bigg) \ ,
\end{align}
valid for $\text{Re}(\lambda) > \tfrac{1}{2}\abs{\text{Re}(\nu)}-1$ and $\text{Re}(\mu)>0$. The remaining integral in \eqref{eq.I2} after the change of variables $m=\sqrt{\tfrac{y}{1-z}}$ is then of the form (Eq.~(7.821.3) of \cite{Gradshteyn:2007})
\begin{align} \label{eq.KGint}
    2\!\int_0^{\infty}\!dy\,y^{-\rho}K_{\nu}\big(2\sqrt{y\,}\big)\left.G^{\,m,n}_{p,q}\,\bigg(\alpha y\,\right|\!\!\!\!\!\!
    \begin{array}{cc}
        &  a_1,\dots,a_p\\
        &  b_1,\dots,b_q
    \end{array}\!\bigg) = \left.G^{\,m,n+2}_{p+2,q}\,\bigg(\alpha\,\right|\!\!\!\!\!\!
    \begin{array}{cc}
        & \rho-\frac{\nu}{2},\rho+\frac{\nu}{2},a_1,\dots,a_p \\
        & b_1,\dots,b_q
    \end{array}\!\bigg) \ ,
\end{align}
with $p+q<2(m+n)$, $\abs{\arg{\alpha}}<(m+n-\tfrac{1}{2}(p+q))\pi$ and $\text{Re}(\rho)<1-\tfrac{1}{2}\text{Re}(\nu) + \min{\text{Re}}(b_j)$.
The Meijer G-function is a particularly general function, designed to include as special cases most other common special functions -- such as the generalised hypergeometric function. The primary definition of this function is in terms of the Mellin-Barnes type line integral
\begin{align} \label{eq.MeijerGdef}
    \left.G^{\,m,n}_{p,q}\,\bigg( x\,\right|\!\!\!\!\!\!
    \begin{array}{cc}
        &  a_1,\dots,a_p\\
        &  b_1,\dots,b_q
    \end{array}\!\bigg) = \frac{1}{2\pi i}\int_L ds\,x^s\,\frac{\prod_{j=1}^{m}\Gamma(b_j-s)\prod_{j=1}^{n}\Gamma(1-a_j+s)}{\prod_{j=m+1}^{q}\Gamma(1-b_j+s)\prod_{j=n+1}^{p}\Gamma(a_j-s)} \ ,
\end{align}
where the choices of integration path $L$ are given in section 9.302 of \cite{Gradshteyn:2007}. Using \eqref{eq.KGint} in \eqref{eq.I2} gives
\begin{align} \label{eq.I2final}
    I_2&= \frac{1}{8}\abs{1-z}^{-1}(1-z)^{-2-\frac{1}{2}(a_1+a_2)}\!\int_0^{\infty}\!dy\, y^{\frac{1}{2}(a_1+a_2+3)}K_1\big(2\sqrt{y}\big) \left.G^{\,2,1}_{1,3}\,\bigg(y\,\frac{1-\zb}{1-z}\,\right| \!\!\!\!\!\!
    \begin{array}{cc}
        & -\frac{a_2}{2} \\
        & 1,0,-\frac{a_2}{2}-1
    \end{array}\!\bigg) \nonumber\\
    &= \frac{1}{16}\abs{1-z}^{-1}(1-z)^{-2-\frac{1}{2}(a_1+a_2)} \left.G^{\,2,3}_{3,3}\,\bigg(\frac{1-\zb}{1-z}\,\right|\!\!\!\!\!\!
    \begin{array}{cc}
        & -\tfrac{1}{2}(a_1+a_2)-2,-\tfrac{1}{2}(a_1+a_2)-1,-\frac{a_2}{2} \\
        & 1,0,-\frac{a_2}{2}-1
    \end{array}\!\bigg)\nonumber\\
    &\approx \frac{1}{16} \eta^{-\frac{1}{2}}\sigma^{-3-\frac{1}{2}(a_1+a_2)}\left.G^{\,2,3}_{3,3}\,\bigg(\eta\;\right|\!\!\!\!\!\!\!
    \begin{array}{cc}
        & -\tfrac{1}{2}(a_1+a_2)-2,-\tfrac{1}{2}(a_1+a_2)-1,-\frac{a_2}{2} \\
        & 1,0,-\frac{a_2}{2}-1
    \end{array}\!\bigg)\nonumber\\
    &\approx \frac{1}{16} \eta^{-\frac{1}{2}} \sigma^{-3-\frac{1}{2}(a_1+a_2)} \Big(1+\frac{a_2}{2}\Big)^{-1}\Gamma\big(2+\tfrac{1}{2}(a_1+a_2)\big)\,\Gamma\big(3+\tfrac{1}{2}(a_1+a_2)\big) \ ,
\end{align}
where in the third and fourth lines an expansion in small $\sigma$ and then $\eta$ is performed. Likewise, $I_1$ can be found by exchanging $z$ and $\zb$ in the second line of \eqref{eq.I2final}, giving
\begin{align} \label{eq.I1}
    I_1 &= \frac{1}{16}\abs{1-z}^{-1}(1-\zb)^{-2-\frac{1}{2}(a_1+a_2)} \left.G^{\,2,3}_{3,3}\,\bigg(\frac{1-z}{1-\zb}\,\right|\!\!\!\!\!\!
    \begin{array}{cc}
        & -\tfrac{1}{2}(a_1+a_2)-2,-\tfrac{1}{2}(a_1+a_2)-1,-\frac{a_2}{2} \\
        & 1,0,-\frac{a_2}{2}-1
    \end{array}\!\bigg) \nonumber\\
    &\approx \frac{1}{16}\sigma^{-3-\frac{1}{2}(a_1+a_2)}\eta^{-\frac{1}{2}(a_1+a_2+5)} \left.G^{\,2,3}_{3,3}\,\bigg(\frac{1}{\eta}\,\right|\!\!\!\!\!\!
    \begin{array}{cc}
        & -\tfrac{1}{2}(a_1+a_2)-2,-\tfrac{1}{2}(a_1+a_2)-1,-\frac{a_2}{2} \\
        & 1,0,-\frac{a_2}{2}-1
    \end{array}\!\bigg) \nonumber\\
    &\approx \frac{1}{16}\sigma^{-3-\frac{1}{2}(a_1+a_2)}\eta^{-\frac{1}{2}(a_1+3)} \,\Gamma\big(1+\frac{a_1}{2}\big)\,\Gamma\big(2+\frac{a_1}{2}\big)\,\Gamma\big(1+\frac{a_2}{2}\big)\,\Gamma\big(2+\frac{a_2}{2}\big) \ .
\end{align}
Using \eqref{eq.I2final} and \eqref{eq.I1} in \eqref{eq.ImReggeCrossingL3} gives
\begin{align} \label{eq.ImReggeCrossingL5}
    \text{Im}\left. C^{\text{bos}}_2\right|_{\lcirclearrowright} &\approx -A\,\pi \sigma^{-4-\frac{1}{2}(a_1+a_2)}\eta^{-1}\bigg[ \Big(1+\frac{a_2}{2}\Big)^{-1}\Gamma\big(2+\tfrac{1}{2}(a_1+a_2)\big)\,\Gamma\big(3+\tfrac{1}{2}(a_1+a_2)\big) \nonumber\\
    &\qquad\qquad\qquad + \eta^{-\frac{a_1}{2}-1} \Gamma\big(1+\frac{a_1}{2}\big)\,\Gamma\big(2+\frac{a_1}{2}\big)\,\Gamma\big(1+\frac{a_2}{2}\big)\,\Gamma\big(2+\frac{a_2}{2}\big)\bigg] \ .
\end{align}
In order to match the stress tensor's leading $\sigma$ and $\eta$ behaviour, it is necessary to have $a_1=0$, $a_2=2$ and $A=-1$.
Plugging these values back into the leading $\sigma$ term of \eqref{eq.ImReggeCrossingL3} and using the following functional relations of the Meijer G-function (see section 9.31 of \cite{Gradshteyn:2007})
\begin{align} \label{eq.Gids}
    \left.G^{\,m,n}_{p,q}\,\bigg(x\,\right|\!\!\!\!\!\!
    \begin{array}{cc}
        & a_1,a_2,\dots,a_p \\
        & b_1,\dots,b_{q-1},a_1
    \end{array}\!\bigg)
    &= \left.G^{\,m,n-1}_{p-1,q-1}\,\bigg(x\,\right|\!\!\!\!\!\!
    \begin{array}{cc}
        & a_2,\dots,a_p \\
        & b_1,\dots,b_{q-1}
    \end{array}\!\bigg) \\
    \left.G^{\,m,n}_{p,q}\,\bigg(x\,\right|\!\!\!\!\!\!
    \begin{array}{cc}
        & a_1,\dots,a_p \\
        & b_1,\dots,b_q
    \end{array}\!\bigg)
    &= x^k \left.G^{\,m,n}_{p,q}\,\bigg(x\,\right|\!\!\!\!\!\!
    \begin{array}{cc}
        & a_1-k,\dots,a_p-k \\
        & b_1-k,\dots,b_q-k
    \end{array}\!\bigg) \ ,
\end{align}
gives
\begin{align} \label{eq.LLLLCrossMeijer}
    \text{Im}\left. C^{\text{bos}}_2\right|_{\lcirclearrowright} &\approx \frac{\pi}{\eta^4\sigma^5}\bigg[\! \left. G^{\,2,2}_{2,2}\,\bigg(\eta^{-1}\,\right|\!\!\!\!\!\!
    \begin{array}{cc}
        & -3,-1\\
        & 1,0
    \end{array}\!\bigg)
    + \left.G^{\,2,2}_{2,2}\,\bigg(\eta\,\right|\!\!\!\!\!\!
    \begin{array}{cc}
        & 0,2\\
        & 4,3
    \end{array}\bigg)\,\bigg] + O(\sigma^{-4}) \ .
\end{align}
The relation
\begin{align}
    \left. G^{\,2,2}_{2,2}\,\bigg(\eta\,\right|\!\!\!\!\!\!
    \begin{array}{cc}
        & a_1,a_2\\
        & b_1,b_2
    \end{array}\!\bigg)
    &= B(1-a_1+b_1,1-a_2+b_2)\,\Gamma(1-a_2+b_1)\,\Gamma(1-a_1+b_2)\,\eta^{b_1} \nonumber\\
    &\qquad\ \ \times\ {}_2F_1\big(1-a_1+b_1,1-a_2+b_1;2-a_1-a_2+b_1+b_2;1-\eta \big) \ ,
\end{align}
with $B(x,y)$ the Euler beta function, allows the Meijer G-functions in \eqref{eq.LLLLCrossMeijer} to be evaluated, giving precisely the leading small $\sigma$ term of the correlator in \eqref{eq.DirectChannelPole100}.

\providecommand{\href}[2]{#2}\begingroup\raggedright\endgroup


\begin{thebibliography}{10}

\bibitem{Amati:1987wq}
D.~Amati, M.~Ciafaloni, and G.~Veneziano, ``{Superstring Collisions at
  Planckian Energies},'' {\em Phys. Lett.} {\bf B197} (1987)
81.

\bibitem{Amati:1987uf}
D.~Amati, M.~Ciafaloni, and G.~Veneziano, ``{Classical and Quantum Gravity
  Effects from Planckian Energy Superstring Collisions},'' {\em Int. J. Mod.
  Phys.} {\bf A3} (1988)
1615--1661.

\bibitem{'tHooft:1987rb}
G.~'t~Hooft, ``{Graviton Dominance in Ultrahigh-Energy Scattering},'' {\em
  Phys. Lett.} {\bf B198} (1987)
61--63.

\bibitem{Cornalba:2006xk}
L.~Cornalba, M.~S. Costa, J.~Penedones, and R.~Schiappa, ``{Eikonal
  Approximation in AdS/CFT: From Shock Waves to Four-Point Functions},'' {\em
  JHEP} {\bf 08} (2007) 019,
\href{http://arXiv.org/abs/hep-th/0611122}{{\tt hep-th/0611122}}.

\bibitem{Cornalba:2006xm}
L.~Cornalba, M.~S. Costa, J.~Penedones, and R.~Schiappa, ``{Eikonal
  Approximation in AdS/CFT: Conformal Partial Waves and Finite N Four-Point
  Functions},'' {\em Nucl. Phys.} {\bf B767} (2007) 327--351,
\href{http://arXiv.org/abs/hep-th/0611123}{{\tt hep-th/0611123}}.

\bibitem{Cornalba:2007zb}
L.~Cornalba, M.~S. Costa, and J.~Penedones, ``{Eikonal approximation in
  AdS/CFT: Resumming the gravitational loop expansion},'' {\em JHEP} {\bf 09}
  (2007) 037,
\href{http://arXiv.org/abs/0707.0120}{{\tt 0707.0120}}.

\bibitem{Cornalba:2007fs}
L.~Cornalba, ``{Eikonal methods in AdS/CFT: Regge theory and multi-reggeon
  exchange},'' \href{http://arXiv.org/abs/0710.5480}{{\tt 0710.5480}}.

\bibitem{Cornalba:2009ax}
L.~Cornalba, M.~S. Costa, and J.~Penedones, ``{Deep Inelastic Scattering in
  Conformal QCD},'' {\em JHEP} {\bf 03} (2010) 133,
  \href{http://arXiv.org/abs/0911.0043}{{\tt 0911.0043}}.

\bibitem{Costa:2012cb}
M.~S. Costa, V.~Goncalves, and J.~Penedones, ``{Conformal Regge theory},'' {\em
  JHEP} {\bf 12} (2012) 091, \href{http://arXiv.org/abs/1209.4355}{{\tt
  1209.4355}}.

\bibitem{Costa:2017twz}
M.~S. Costa, T.~Hansen, and J.~Penedones, ``{Bounds for OPE coefficients on the
  Regge trajectory},'' {\em JHEP} {\bf 10} (2017) 197,
  \href{http://arXiv.org/abs/1707.07689}{{\tt 1707.07689}}.

\bibitem{Kulaxizi:2017ixa}
M.~Kulaxizi, A.~Parnachev, and A.~Zhiboedov, ``{Bulk Phase Shift, CFT Regge
  Limit and Einstein Gravity},'' {\em JHEP} {\bf 06} (2018) 121,
  \href{http://arXiv.org/abs/1705.02934}{{\tt 1705.02934}}.

\bibitem{Li:2017lmh}
D.~Li, D.~Meltzer, and D.~Poland, ``{Conformal Bootstrap in the Regge Limit},''
  {\em JHEP} {\bf 12} (2017) 013,
\href{http://arXiv.org/abs/1705.03453}{{\tt 1705.03453}}.

\bibitem{D'Appollonio:2010ae}
G.~D'Appollonio, P.~Di~Vecchia, R.~Russo, and G.~Veneziano, ``{High-energy
  string-brane scattering: leading eikonal and beyond},'' {\em JHEP} {\bf 11}
  (2010) 100,
\href{http://arXiv.org/abs/1008.4773}{{\tt 1008.4773}}.

\bibitem{Kulaxizi:2018dxo}
M.~Kulaxizi, G.~S. Ng, and A.~Parnachev, ``{Black Holes, Heavy States, Phase
  Shift and Anomalous Dimensions},'' {\em SciPost Phys.} {\bf 6} (2019), no.~6,
  065,
\href{http://arXiv.org/abs/1812.03120}{{\tt 1812.03120}}.

\bibitem{Karlsson:2019qfi}
R.~Karlsson, M.~Kulaxizi, A.~Parnachev, and P.~Tadi\'c, ``{Black Holes and
  Conformal Regge Bootstrap},'' {\em JHEP} {\bf 10} (2019) 046,
  \href{http://arXiv.org/abs/1904.00060}{{\tt 1904.00060}}.

\bibitem{Strominger:1996sh}
A.~Strominger and C.~Vafa, ``{Microscopic origin of the Bekenstein-Hawking
  entropy},'' {\em Phys.Lett.} {\bf B379} (1996) 99--104,
  \href{http://arXiv.org/abs/hep-th/9601029}{{\tt hep-th/9601029}}.

\bibitem{Maldacena:1997re}
J.~M. Maldacena, ``{The large N limit of superconformal field theories and
  supergravity},'' {\em Adv. Theor. Math. Phys.} {\bf 2} (1998) 231--252,
\href{http://arXiv.org/abs/hep-th/9711200}{{\tt hep-th/9711200}}.

\bibitem{Lunin:2001jy}
O.~Lunin and S.~D. Mathur, ``{AdS/CFT duality and the black hole information
  paradox},'' {\em Nucl. Phys.} {\bf B623} (2002) 342--394,
\href{http://arXiv.org/abs/hep-th/0109154}{{\tt hep-th/0109154}}.

\bibitem{Bena:2006kb}
I.~Bena, C.-W. Wang, and N.~P. Warner, ``{Mergers and Typical Black Hole
  Microstates},'' {\em JHEP} {\bf 11} (2006) 042,
\href{http://arXiv.org/abs/hep-th/0608217}{{\tt hep-th/0608217}}.

\bibitem{Kanitscheider:2007wq}
I.~Kanitscheider, K.~Skenderis, and M.~Taylor, ``{Fuzzballs with internal
  excitations},'' {\em JHEP} {\bf 06} (2007) 056,
\href{http://arXiv.org/abs/0704.0690}{{\tt 0704.0690}}.

\bibitem{Bena:2016ypk}
I.~Bena, S.~Giusto, E.~J. Martinec, R.~Russo, M.~Shigemori, D.~Turton, and
  N.~P. Warner, ``{Smooth horizonless geometries deep inside the black-hole
  regime},'' {\em Phys. Rev. Lett.} {\bf 117} (2016), no.~20, 201601,
\href{http://arXiv.org/abs/1607.03908}{{\tt 1607.03908}}.

\bibitem{Warner:2019jll}
N.~P. Warner, ``{Lectures on Microstate Geometries},''
  \href{http://arXiv.org/abs/1912.13108}{{\tt 1912.13108}}.

\bibitem{Mathur:2009hf}
S.~D. Mathur, ``{The information paradox: A pedagogical introduction},'' {\em
  Class. Quant. Grav.} {\bf 26} (2009) 224001,
\href{http://arXiv.org/abs/0909.1038}{{\tt 0909.1038}}.

\bibitem{Almheiri:2012rt}
A.~Almheiri, D.~Marolf, J.~Polchinski, and J.~Sully, ``{Black Holes:
  Complementarity or Firewalls?},'' {\em JHEP} {\bf 02} (2013) 062,
\href{http://arXiv.org/abs/1207.3123}{{\tt 1207.3123}}.

\bibitem{Bena:2015bea}
I.~Bena, S.~Giusto, R.~Russo, M.~Shigemori, and N.~P. Warner, ``{Habemus
  Superstratum! A constructive proof of the existence of superstrata},'' {\em
  JHEP} {\bf 1505} (2015) 110,
\href{http://arXiv.org/abs/1503.01463}{{\tt 1503.01463}}.

\bibitem{Bena:2017xbt}
I.~Bena, S.~Giusto, E.~J. Martinec, R.~Russo, M.~Shigemori, D.~Turton, and
  N.~P. Warner, ``{Asymptotically-flat supergravity solutions deep inside the
  black-hole regime},'' {\em JHEP} {\bf 02} (2018) 014,
  \href{http://arXiv.org/abs/1711.10474}{{\tt 1711.10474}}.

\bibitem{Bena:2019azk}
I.~Bena, P.~Heidmann, R.~Monten, and N.~P. Warner, ``{Thermal Decay without
  Information Loss in Horizonless Microstate Geometries},'' {\em SciPost Phys.}
  {\bf 7} (2019), no.~5, 063, \href{http://arXiv.org/abs/1905.05194}{{\tt
  1905.05194}}.

\bibitem{Bena:2020yii}
I.~Bena, F.~Eperon, P.~Heidmann, and N.~P. Warner, ``{The Great Escape:
  Tunneling out of Microstate Geometries},''
  \href{http://arXiv.org/abs/2005.11323}{{\tt 2005.11323}}.

\bibitem{Galliani:2016cai}
A.~Galliani, S.~Giusto, E.~Moscato, and R.~Russo, ``{Correlators at large c
  without information loss},'' {\em JHEP} {\bf 09} (2016) 065,
\href{http://arXiv.org/abs/1606.01119}{{\tt 1606.01119}}.

\bibitem{Galliani:2017jlg}
A.~Galliani, S.~Giusto, and R.~Russo, ``{Holographic 4-point correlators with
  heavy states},'' {\em JHEP} {\bf 10} (2017) 040,
\href{http://arXiv.org/abs/1705.09250}{{\tt 1705.09250}}.

\bibitem{Bombini:2017sge}
A.~Bombini, A.~Galliani, S.~Giusto, E.~Moscato, and R.~Russo, ``{Unitary
  4-point correlators from classical geometries},'' {\em Eur. Phys. J.} {\bf
  C78} (2018), no.~1, 8,
\href{http://arXiv.org/abs/1710.06820}{{\tt 1710.06820}}.

\bibitem{Bombini:2019vnc}
A.~Bombini and A.~Galliani, ``{AdS$_{3}$ four-point functions from $
  \frac{1}{8} $ -BPS states},'' {\em JHEP} {\bf 06} (2019) 044,
  \href{http://arXiv.org/abs/1904.02656}{{\tt 1904.02656}}.

\bibitem{Tian:2019ash}
J.~Tian, J.~Hou, and B.~Chen, ``{Holographic Correlators on Integrable
  Superstrata},'' {\em Nucl. Phys. B} {\bf 948} (2019) 114766,
  \href{http://arXiv.org/abs/1904.04532}{{\tt 1904.04532}}.

\bibitem{Tyukov:2017uig}
A.~Tyukov, R.~Walker, and N.~P. Warner, ``{Tidal Stresses and Energy Gaps in
  Microstate Geometries},'' {\em JHEP} {\bf 02} (2018) 122,
  \href{http://arXiv.org/abs/1710.09006}{{\tt 1710.09006}}.

\bibitem{Bianchi:2017sds}
M.~Bianchi, D.~Consoli, and J.~Morales, ``{Probing Fuzzballs with Particles,
  Waves and Strings},'' {\em JHEP} {\bf 06} (2018) 157,
  \href{http://arXiv.org/abs/1711.10287}{{\tt 1711.10287}}.

\bibitem{Bianchi:2018kzy}
M.~Bianchi, D.~Consoli, A.~Grillo, and J.~F. Morales, ``{The dark side of
  fuzzball geometries},'' {\em JHEP} {\bf 05} (2019) 126,
  \href{http://arXiv.org/abs/1811.02397}{{\tt 1811.02397}}.

\bibitem{Bena:2018mpb}
I.~Bena, E.~J. Martinec, R.~Walker, and N.~P. Warner, ``{Early Scrambling and
  Capped BTZ Geometries},'' {\em JHEP} {\bf 04} (2019) 126,
  \href{http://arXiv.org/abs/1812.05110}{{\tt 1812.05110}}.

\bibitem{Bianchi:2020des}
M.~Bianchi, A.~Grillo, and J.~F. Morales, ``{Chaos at the rim of black hole and
  fuzzball shadows},'' {\em JHEP} {\bf 05} (2020) 078,
  \href{http://arXiv.org/abs/2002.05574}{{\tt 2002.05574}}.

\bibitem{Bena:2020iyw}
I.~Bena, A.~Houppe, and N.~P. Warner, ``{Delaying the Inevitable: Tidal
  Disruption in Microstate Geometries},''
  \href{http://arXiv.org/abs/2006.13939}{{\tt 2006.13939}}.

\bibitem{Giusto:2018ovt}
S.~Giusto, R.~Russo, and C.~Wen, ``{Holographic correlators in AdS$_{3}$},''
  {\em JHEP} {\bf 03} (2019) 096,
\href{http://arXiv.org/abs/1812.06479}{{\tt 1812.06479}}.

\bibitem{Rastelli:2019gtj}
L.~Rastelli, K.~Roumpedakis, and X.~Zhou, ``{$\mathbf{AdS_3\times S^3}$
  Tree-Level Correlators: Hidden Six-Dimensional Conformal Symmetry},'' {\em
  JHEP} {\bf 10} (2019) 140, \href{http://arXiv.org/abs/1905.11983}{{\tt
  1905.11983}}.

\bibitem{Giusto:2019pxc}
S.~Giusto, R.~Russo, A.~Tyukov, and C.~Wen, ``{Holographic correlators in
  AdS$_3$ without Witten diagrams},'' {\em JHEP} {\bf 09} (2019) 030,
\href{http://arXiv.org/abs/1905.12314}{{\tt 1905.12314}}.

\bibitem{Giusto:2020neo}
S.~Giusto, R.~Russo, A.~Tyukov, and C.~Wen, ``{The CFT$_6$ origin of all
  tree-level 4-point correlators in AdS$_3 \times S^3$},''
\href{http://arXiv.org/abs/2005.08560}{{\tt 2005.08560}}.

\bibitem{Heemskerk:2009pn}
I.~Heemskerk, J.~Penedones, J.~Polchinski, and J.~Sully, ``{Holography from
  Conformal Field Theory},'' {\em JHEP} {\bf 10} (2009) 079,
\href{http://arXiv.org/abs/0907.0151}{{\tt 0907.0151}}.

\bibitem{Fitzpatrick:2015zha}
A.~L. Fitzpatrick, J.~Kaplan, and M.~T. Walters, ``{Virasoro Conformal Blocks
  and Thermality from Classical Background Fields},'' {\em JHEP} {\bf 11}
  (2015) 200, \href{http://arXiv.org/abs/1501.05315}{{\tt 1501.05315}}.

\bibitem{Fitzpatrick:2014vua}
A.~L. Fitzpatrick, J.~Kaplan, and M.~T. Walters, ``{Universality of
  Long-Distance AdS Physics from the CFT Bootstrap},'' {\em JHEP} {\bf 08}
  (2014) 145,
\href{http://arXiv.org/abs/1403.6829}{{\tt 1403.6829}}.

\bibitem{Balasubramanian:2000rt}
V.~Balasubramanian, J.~de~Boer, E.~Keski-Vakkuri, and S.~F. Ross,
  ``{Supersymmetric conical defects: Towards a string theoretic description of
  black hole formation},'' {\em Phys. Rev.} {\bf D64} (2001) 064011,
\href{http://arXiv.org/abs/hep-th/0011217}{{\tt hep-th/0011217}}.

\bibitem{Maldacena:2000dr}
J.~M. Maldacena and L.~Maoz, ``{Desingularization by rotation},'' {\em JHEP}
  {\bf 0212} (2002) 055,
\href{http://arXiv.org/abs/hep-th/0012025}{{\tt hep-th/0012025}}.

\bibitem{Li:2020dqm}
  Y.~Z.~Li and H.~Y.~Zhang, {``More on Heavy-Light Bootstrap up to Double-Stress-Tensor,''}
\href{http://arXiv.org/abs/2004.04758}{{\tt 2004.04758}}.

\bibitem{Kraus:2018zrn}
P.~Kraus, A.~Sivaramakrishnan and R.~Snively, ``{Late time Wilson lines,}''
{\em JHEP} \textbf{04} (2019), 026,
\href{http://arXiv.org/abs/1810.01439}{{\tt 1810.01439}}.


\bibitem{Alday:2017gde}
L.~F. Alday, A.~Bissi, and E.~Perlmutter, ``{Holographic Reconstruction of AdS
  Exchanges from Crossing Symmetry},'' {\em JHEP} {\bf 08} (2017) 147,
\href{http://arXiv.org/abs/1705.02318}{{\tt 1705.02318}}.

\bibitem{Alday:2017vkk}
L.~F. Alday and S.~Caron-Huot, ``{Gravitational S-matrix from CFT dispersion
  relations},'' {\em JHEP} {\bf 12} (2018) 017,
\href{http://arXiv.org/abs/1711.02031}{{\tt 1711.02031}}.

\bibitem{Caron-Huot:2017vep}
S.~Caron-Huot, ``{Analyticity in Spin in Conformal Theories},'' {\em JHEP} {\bf
  09} (2017) 078,
\href{http://arXiv.org/abs/1703.00278}{{\tt 1703.00278}}.

\bibitem{Gradshteyn:2007}
I.~S. Gradshteyn and I.~M. Ryzhik, {\em {Table of Integrals, Series, and
  Products (Seventh Edition)}}.
\newblock Academic Press, San Diego; London, 2007.

\end{thebibliography}
\end{document}